\documentclass[12pt,titlepage,letterpaper]{utarticle}

\usepackage{amsmath}
\usepackage{amsfonts}
\usepackage{amssymb}
\usepackage{amsthm}
\usepackage{mathtools}
\usepackage{graphicx}
\usepackage{color}
\usepackage{xparse}
\usepackage[utf8x]{inputenc}
\usepackage{tocloft}
\usepackage[titletoc,title]{appendix}
\usepackage{hyperref}
\usepackage{longtable}

\definecolor{aqua}{rgb}{0, 1.0, 1.0}
\definecolor{fuschia}{rgb}{1.0, 0, 1.0}
\definecolor{gray}{rgb}{0.502, 0.502, 0.502}
\definecolor{lime}{rgb}{0, 1.0, 0}
\definecolor{maroon}{rgb}{0.502, 0, 0}
\definecolor{navy}{rgb}{0, 0, 0.502}
\definecolor{olive}{rgb}{0.502, 0.502, 0}
\definecolor{purple}{rgb}{0.502, 0, 0.502}
\definecolor{silver}{rgb}{0.753, 0.753, 0.753}
\definecolor{teal}{rgb}{0, 0.502, 0.502}

\makeatletter
\newdimen\itex@wd%
\newdimen\itex@dp%
\newdimen\itex@thd%
\def\itexspace#1#2#3{\itex@wd=#3em%
\itex@wd=0.1\itex@wd%
\itex@dp=#2ex%
\itex@dp=0.1\itex@dp%
\itex@thd=#1ex%
\itex@thd=0.1\itex@thd%
\advance\itex@thd\the\itex@dp%
\makebox[\the\itex@wd]{\rule[-\the\itex@dp]{0cm}{\the\itex@thd}}}
\makeatother

\makeatletter
\newif\if@sup
\newtoks\@sups
\def\append@sup#1{\edef\act{\noexpand\@sups={\the\@sups #1}}\act}%
\def\reset@sup{\@supfalse\@sups={}}%
\def\mk@scripts#1#2{\if #2/ \if@sup ^{\the\@sups}\fi \else%
  \ifx #1_ \if@sup ^{\the\@sups}\reset@sup \fi {}_{#2}%
  \else \append@sup#2 \@suptrue \fi%
  \expandafter\mk@scripts\fi}
\def\tensor#1#2{\reset@sup#1\mk@scripts#2_/}
\def\multiscripts#1#2#3{\reset@sup{}\mk@scripts#1_/#2%
  \reset@sup\mk@scripts#3_/}
\makeatother

\makeatletter
\newbox\slashbox \setbox\slashbox=\hbox{$/$}
\def\itex@pslash#1{\setbox\@tempboxa=\hbox{$#1$}
  \@tempdima=0.5\wd\slashbox \advance\@tempdima 0.5\wd\@tempboxa
  \copy\slashbox \kern-\@tempdima \box\@tempboxa}
\def\slash{\protect\itex@pslash}
\makeatother

\def\clap#1{\hbox to 0pt{\hss#1\hss}}

\let\oldroot\root
\def\root#1#2{\oldroot #1 \of{#2}}
\renewcommand{\sqrt}[2][]{\oldroot #1 \of{#2}}

\DeclareSymbolFont{symbolsC}{U}{txsyc}{m}{n}
\SetSymbolFont{symbolsC}{bold}{U}{txsyc}{bx}{n}
\DeclareFontSubstitution{U}{txsyc}{m}{n}

\DeclareSymbolFont{stmry}{U}{stmry}{m}{n}
\SetSymbolFont{stmry}{bold}{U}{stmry}{b}{n}

\DeclareFontFamily{OMX}{MnSymbolE}{}
\DeclareSymbolFont{mnomx}{OMX}{MnSymbolE}{m}{n}
\SetSymbolFont{mnomx}{bold}{OMX}{MnSymbolE}{b}{n}
\DeclareFontShape{OMX}{MnSymbolE}{m}{n}{
    <-6>  MnSymbolE5
   <6-7>  MnSymbolE6
   <7-8>  MnSymbolE7
   <8-9>  MnSymbolE8
   <9-10> MnSymbolE9
  <10-12> MnSymbolE10
  <12->   MnSymbolE12}{}

\makeatletter
\def\re@DeclareMathSymbol#1#2#3#4{%
    \let#1=\undefined
    \DeclareMathSymbol{#1}{#2}{#3}{#4}}
\re@DeclareMathSymbol{\neArrow}{\mathrel}{symbolsC}{116}
\re@DeclareMathSymbol{\neArr}{\mathrel}{symbolsC}{116}
\re@DeclareMathSymbol{\seArrow}{\mathrel}{symbolsC}{117}
\re@DeclareMathSymbol{\seArr}{\mathrel}{symbolsC}{117}
\re@DeclareMathSymbol{\nwArrow}{\mathrel}{symbolsC}{118}
\re@DeclareMathSymbol{\nwArr}{\mathrel}{symbolsC}{118}
\re@DeclareMathSymbol{\swArrow}{\mathrel}{symbolsC}{119}
\re@DeclareMathSymbol{\swArr}{\mathrel}{symbolsC}{119}
\re@DeclareMathSymbol{\nequiv}{\mathrel}{symbolsC}{46}
\re@DeclareMathSymbol{\Perp}{\mathrel}{symbolsC}{121}
\re@DeclareMathSymbol{\Vbar}{\mathrel}{symbolsC}{121}
\re@DeclareMathSymbol{\sslash}{\mathrel}{stmry}{12}
\re@DeclareMathSymbol{\bigsqcap}{\mathop}{stmry}{"64}
\re@DeclareMathSymbol{\biginterleave}{\mathop}{stmry}{"6}
\re@DeclareMathSymbol{\invamp}{\mathrel}{symbolsC}{77}
\re@DeclareMathSymbol{\parr}{\mathrel}{symbolsC}{77}
\makeatother

\makeatletter
\def\Decl@Mn@Delim#1#2#3#4{%
  \if\relax\noexpand#1%
    \let#1\undefined
  \fi
  \DeclareMathDelimiter{#1}{#2}{#3}{#4}{#3}{#4}}
\def\Decl@Mn@Open#1#2#3{\Decl@Mn@Delim{#1}{\mathopen}{#2}{#3}}
\def\Decl@Mn@Close#1#2#3{\Decl@Mn@Delim{#1}{\mathclose}{#2}{#3}}
\Decl@Mn@Open{\llangle}{mnomx}{'164}
\Decl@Mn@Close{\rrangle}{mnomx}{'171}
\Decl@Mn@Open{\lmoustache}{mnomx}{'245}
\Decl@Mn@Close{\rmoustache}{mnomx}{'244}
\makeatother

\makeatletter
\DeclareRobustCommand\widecheck[1]{{\mathpalette\@widecheck{#1}}}
\def\@widecheck#1#2{%
    \setbox\z@\hbox{\m@th$#1#2$}%
    \setbox\tw@\hbox{\m@th$#1%
       \widehat{%
          \vrule\@width\z@\@height\ht\z@
          \vrule\@height\z@\@width\wd\z@}$}%
    \dp\tw@-\ht\z@
    \@tempdima\ht\z@ \advance\@tempdima2\ht\tw@ \divide\@tempdima\thr@@
    \setbox\tw@\hbox{%
       \raise\@tempdima\hbox{\scalebox{1}[-1]{\lower\@tempdima\box
\tw@}}}%
    {\ooalign{\box\tw@ \cr \box\z@}}}
\makeatother

\makeatletter
\NewDocumentCommand\mathraisebox{moom}{%
\IfNoValueTF{#2}{\def\@temp##1##2{\raisebox{#1}{$\m@th##1##2$}}}{%
\IfNoValueTF{#3}{\def\@temp##1##2{\raisebox{#1}[#2]{$\m@th##1##2$}}%
}{\def\@temp##1##2{\raisebox{#1}[#2][#3]{$\m@th##1##2$}}}}%
\mathpalette\@temp{#4}}
\makeatletter

\makeatletter
\def\udots{\mathinner{\mkern2mu\raise\p@\hbox{.}
\mkern2mu\raise4\p@\hbox{.}\mkern1mu
\raise7\p@\vbox{\kern7\p@\hbox{.}}\mkern1mu}}
\makeatother

\newcommand{\gt}{>}

\theoremstyle{plain}

\theoremstyle{definition}

\theoremstyle{remark}

\numberwithin{equation}{section}

\def\Spin{\mathrm{Spin}}

\allowdisplaybreaks



\begin{document}


\preprint{
UTTG--12--14\\
TCC--012--14\\
ICTP--SAIFR/2014--002\\
}

\title{Seiberg-Witten for $Spin(n)$ with Spinors}

\author{Oscar Chacaltana
    \address{
    ICTP South American Institute for\\ Fundamental Research,\\
    Instituto de F\'isica Te\'orica,\\Universidade Estadual Paulista,\\
    01140-070 S\~{a}o Paulo, SP, Brazil\\
    {~}\\
    \email{chacaltana@ift.unesp.br}\\
    },
    Jacques Distler ${}^\mathrm{b}$ and Anderson Trimm
     \address{
      Theory Group and\\
      Texas Cosmology Center\\
      Department of Physics,\\
      University of Texas at Austin,\\
      Austin, TX 78712, USA \\
      {~}\\
      \email{distler@golem.ph.utexas.edu}\\
      \email{atrimm@physics.utexas.edu}
      }
}
\date{April 14, 2014}

\Abstract{
$\mathcal{N}=2$ supersymmetric $Spin(n)$ gauge theory admits hypermultiplets in spinor representations of the gauge group, compatible with $\beta\leq0$, for $n\leq 14$. The theories with $\beta<0$ can be obtained as mass-deformations of the $\beta=0$ theories, so it is of greatest interest to construct the $\beta=0$ theories. In previous works, we discussed the $n\leq8$ theories. Here, we turn to the $9\leq n\leq 14$ cases. By compactifying the $D_N$ (2,0) theory on a 4-punctured sphere, we find Seiberg-Witten solutions to almost all of the remaining cases. There are five theories, however, which do not seem to admit a realization from six dimensions.
}

\maketitle

\tocloftpagestyle{empty}
\tableofcontents
\vfill
\newpage
\setcounter{page}{1}

\section{Introduction}\label{introduction}

$\mathcal{N}=2$ supersymmetric $Spin(n)$ gauge theory, with $n-2$ hypermultiplets in the vector representation, is superconformal for any $n\gt2$, and the Seiberg-Witten solutions are known from the mid 1990's \cite{Argyres:1995fw,Hanany:1995fu}. Replacing some number of vectors by hypermultiplets in spinor representations is only possible for sufficiently low $n$. The corresponding Seiberg-Witten solutions do not seem to be known\footnote{The solutions (with arbitrary masses for the vector and spinor hypermultiplets) of the asymptotically-free theories for $n=8,10,12$ were constructed in \cite{Terashima:1998fx}. The status of Seiberg-Witten solutions, to \emph{various} $\mathcal{N}=2$ supersymmetric gauge theories, was recently reviewed in \cite{Bhardwaj:2013qia}.}. For $Spin(5)\simeq Sp(2)$ and $\Spin(6)\simeq SU(4)$, the solutions were presented in \cite{Chacaltana:2010ks,Chacaltana:2012ch}. The solutions to $Spin(7),Spin(8)$ appeared in our previous papers \cite{Chacaltana:2011ze,Chacaltana:2013oka} (see \cite{Tachikawa:2011yr} for an alternative formulation). As a further application of \cite{Chacaltana:2011ze,Chacaltana:2013oka}, we will discuss $Spin(n)$ gauge theories for $n=9,10,\dots,14$, with matter content such that $\beta=0$. These are all of the remaining cases where one can have matter in the spinor representation. For $n\gt14$, only matter in the vector representation is compatible with $\beta\leq0$.

These 4D gauge theories can be obtained by compactifying \cite{Gaiotto:2009we,Gaiotto:2009hg} a 6D (2,0) theory of type $D_N$ on a 4-punctured sphere, where the punctures are labeled by nilpotent orbits in $\mathfrak{d}_N$ (or in $\mathfrak{c}_{N-1}$ for twisted-sector punctures) \cite{Tachikawa:2009rb,Tachikawa:2010vg,Chacaltana:2011ze,Chacaltana:2013oka,Chacaltana:2012zy}. When the 4-punctured sphere degenerates into a pair of 3-punctured spheres (``fixtures"), connected by a long thin cylinder, the gauge theory description is weakly-coupled. Fixtures with only hypermultiplets in the vector representation are, necessarily, twisted. With at least one (half-)hypermultiplet in the spinor representation, we can find an untwisted fixture and --- wherever possible --- we prefer to work in the untwisted theory.

From these realizations as 4-punctured spheres, we construct the corresponding Seiberg-Witten geometries, and discuss the strong-coupling S-dual realizations \cite{Argyres:2007cn} of the gauge theories.

\section{Seiberg-Witten Geometry}\label{SeibergWittenGeometry}
\subsection{Seiberg-Witten curve}\label{SeibergWittenCurve}

In the $D_N$ theory, the Seiberg-Witten curve, $\Sigma\subset \text{tot}(K_C)$, is the spectral curve (in the vector representation) for $D_N$. In other words, it can be written as the locus
\begin{equation}\label{SWDcurve}
  0 = \lambda^{2N} + \phi_2(z) \lambda^{2N-2} + \phi_4(z)\lambda^{2N-4}+\dots+\phi_{2N-2}(z) \lambda^2 + \tilde{\phi}(z)^2
\end{equation}
where the Seiberg-Witten differential, $\lambda=ydz$, is the tautological 1-form on $K_C$. $\Sigma$ is a branched cover of $C$,  of rather high genus. But it admits an obvious involution $\iota\colon\lambda\to-\lambda$. The quotient by this involution is a curve $\tilde{C}$, also a branched cover of $C$. One finds\footnote{For many purposes, it's convenient to replace $\Sigma$ by the compact curve
$$
  0 = \lambda^{2N} + \phi_2(z) \lambda^{2N-2}\mu^2 + \phi_4(z)\lambda^{2N-4}\mu^4+\dots+\phi_{2N-2}(z) \lambda^2\mu^{2N-2} + \tilde{\phi}(z)^2 \mu^{2N}
$$
in $\text{tot}(P(K_C\oplus \mathcal{O}))$.  Away from the punctures, $\mu\neq 0$ and we can scale it to $1$. At the punctures, $\mu=0$, and the SW curve has interesting ramification over the punctures. The $A_{N-1}$ case \cite{MehtaSeshadri,Simpson_HarmonicBundles} is explained in detail in \cite{Chuang:2013wpa}. The generalization to $D_N$ has a few subtleties, which we won't attempt to explicate here.
} that $g(\Sigma)-g(\tilde{C})=N$. The SW solution is obtained by computing the periods of $\lambda$ over the cycles which are anti-invariant under $\iota$. Said differently, the fibers of the Hitchin integrable system are the Prym variety for $\Sigma\to\tilde{C}$.

For the $Spin(2N)$ gauge theories considered below, the above description is completely adequate, as $\tilde{\phi}(z)$ is nowhere-vanishing on $C$. For the $Spin(2N-1)$ gauge theories, $\tilde{\phi}(z)$ vanishes identically. So $\Sigma$ is reducible
$$
   0 = \lambda^2( \lambda^{2N-2} + \phi_2(z) \lambda^{2N-4} + \phi_4(z)\lambda^{2N-6}+\dots+\phi_{2N-2}(z))\quad.
$$
Let $\Sigma_0$ be the component
$$
   0 = \lambda^{2N-2} + \phi_2(z) \lambda^{2N-4} + \phi_4(z)\lambda^{2N-6}+\dots+\phi_{2N-2}(z)\quad.
$$
As before, $\Sigma_0$ admits an involution $\iota\colon\lambda\to-\lambda$, with quotient $\tilde{C}_0 = \Sigma_0/\iota$, and the SW solution, for the $Spin(2N-1)$ gauge theory, is given by the periods of $\lambda$ on the anti-invariant cycles. There is one subtlety which did not occur in the previous case: $\phi_{2N-2}(z)$ typically does have zeroes on $C$, which means that  $\Sigma_0$  is slightly singular. It has ordinary double-points over the zeroes of $\phi_{2N-2}(z)$. As in Hitchin's original paper \cite{Hitchin_StableBundles}, we actually work over the resolutions\footnote{In the $D_4$ theory, there are examples of $Spin(8)$ gauge theory, with matter in the $n_s(8_s)+n_c(8_c)+ (6-n_s-n_c)(8_v)$, where $\tilde{\phi}(z)$ has isolated zeroes on $C$. Over those points, $\Sigma$ has ordinary double points and, similarly, we work on the resolution, $\hat{\Sigma}$.}, $\hat{\Sigma}_0\to \tilde{C}_0$, whose Prym variety has the desired dimension, $g(\hat{\Sigma}_0)-g(\tilde{C}_0)= N-1$.

\subsection{Calabi-Yau geometry}\label{CalabiYauGeometry}
An alternative formulation \cite{Klemm:1996bj,Diaconescu:2006ry}, more directly related to the Type-IIB description of these 4D theories is as follows. Consider a family of noncompact Calabi-Yau 3-folds, $X_{\vec{u}}$, realized as the hypersurface 
\[\label{CYhypersurface}
0 = w^2 + y x^2 - y^{N-1} - \phi_2(z) y^{N-2} - \phi_4(z) y^{N-3} -\dots - \phi_{2N-2}(z) -2 \tilde{\phi}(z) x
\]
in the total space of the bundle $V=\bigl(K_C^{(N-1)}\oplus K_C^{(N-2)}\oplus K_C^2\bigr)\to C$. Here, $\vec{u}$ are the Coulomb branch parameters, on which the $\phi_k(z)$ depend, and
$$w=\tilde{w}(dz)^{N-1},\quad x=\tilde{x}(dz)^{N-2},\quad y=\tilde{y}(dz)^2
$$
are the tautological differentials on $V$. The $g_s\to 0$ limit of Type IIB on $\mathbb{R}^{3,1}\times X_{\vec{u}}$ is the $4D$ $\mathcal{N}=2$ field theory (decoupled from the bulk gravity).

$X_{\vec{u}}$ has a collection of 3-cycles of the form of an $S^2$ in the fiber over a curve on $C$. The Seiberg-Witten solutions to the $Spin(2N)$ theories below are constructed from the periods of the holomorphic 3-form,
$$
    \Omega = \frac{d\tilde{x}\wedge d\tilde{y}\wedge dz}{\tilde{w}}
$$
over a (rational) symplectic basis of these 3-cycles. For the $Spin(2N-1)$ theories, $\tilde{\phi}(z)\equiv 0$, and $X_{\vec{u}}$ has an involution $\iota\colon (w,x) \to (-w,-x)$, under which $\Omega$ is invariant. $\iota$ acts by exchanging two of the $S^2$s in the fiber (fixing the rest). Integrating $\Omega$ over the invariant cycles yields the  $2(N-1)$ periods which comprise the solution for the $Spin(2N-1)$ theories.

\subsection{Dependence on the gauge coupling}\label{dependence}
The Seiberg-Witten solutions to the $\beta=0$ gauge theories, which are our focus, have elaborate (but holomorphic) dependence \cite{Seiberg:1994aj} on the complexified gauge coupling
$$
\tau = \frac{\theta}{\pi} +\frac{8\pi i}{g^2}\quad.
$$
In particular, \emph{any} such theory, which can be realized by compactifying the (2,0) theory on a 4-punctured sphere, \emph{automatically} has a symmetry under $\Gamma(2)\subset PSL(2,\mathbb{Z})$, generated by
\begin{displaymath}
T^2:\,\tau\mapsto \tau+2,\quad S T^2 S:\,\tau\mapsto \frac{\tau}{1-2\tau}\quad.
\end{displaymath}
That is, the dependence on the gauge coupling is through the function
\begin{displaymath}
\begin{split}
   f(\tau) &\equiv -\frac{\theta_2^4(0,\tau)}{\theta_4^4(0,\tau)}\\
           &= -\left(16 q^{1/2} + 128 q +704 q^{3/2}+\dots\right)
\end{split}
\end{displaymath}
where $q=e^{2\pi i \tau}$. 

In the untwisted theory, $f(\tau)$ is simply identified with the cross-ratio of the 4-punctured sphere:
\begin{equation}\label{untwistedCoupling}
f(\tau)= x\equiv \frac{z_{1 3} z_{2 4}}{z_{1 4} z_{2 3}}\quad.
\end{equation}
The limit $x\to 0$ is the usual weak-coupling limit. $x\to 1$ and $x\to\infty$ are limits which admit an  alternative (physically-distinct) S-dual description as a weakly coupled gauge theory.

When the punctures at $z_1$ and $z_2$ are identical, then the theory has a larger symmetry under $\Gamma_0(2)\supset\Gamma(2)$, where the extra generator acts on the $x$-plane as
$$
S: x\mapsto \frac{1}{x}\quad.
$$
The theories, below, with two (one full and one minimal) twisted punctures and two untwisted punctures, have a similar story, except that the relation between $f(\tau)$ (which parametrizes the gauge theory moduli space) and the cross-ratio is more complicated. The gauge theory moduli space is a branched double-cover \cite{Chacaltana:2013oka} of the moduli space of the 4-punctured sphere, $\mathcal{M}_{0,4}$. Instead of \eqref{untwistedCoupling},
\begin{equation}
w^2 = x \equiv \frac{z_{1 3} z_{2 4}}{z_{1 4} z_{2 3}}
\end{equation}
and the gauge coupling
\begin{equation}\label{twistedCoupling}
f(\tau) = \frac{w-1}{w+1}\quad.
\end{equation}
In particular, this means that $x\to 0$ corresponds to $f(\tau)\to -1$ (i.e.~$\tau\to i$), which is an \emph{interior} point of the gauge theory moduli space and intrinsically strongly coupled. As in our previous works on the twisted sector \cite{Chacaltana:2012ch,Chacaltana:2013oka}, we denote these peculiar degenerations as involving a ``gauge theory fixture." The other degeneration limits have more prosaic interpretations. The limit $f(\tau)\to 1$ projects to $x\to\infty$ and the limits $f(\tau)\to 0$ and $f(\tau)\to\infty$ (which have isomorphic physics) both project to $x\to 1$.

In presenting the solutions, below, we write the dependence on the positions of the four punctures in a manifestly $PSL(2,\mathbb{C})$-invariant form. For calculational purposes, it is invariably easier to fix the $PSL(2,\mathbb{C})$ symmetry by setting $(z_1,z_2,z_3,z_4)=(0,\infty,x,1)$.

\section{$Spin(2N) + (2N-2)(V)$ and $Spin(2N-1) + (2N-3)(V)$}\label{Spin2N_and_Spin2Nm1}

Just as $Spin(2N)$ gauge theory with $2(N-1)$ fundamentals is realized as the compactification of the $D_N$ theory with four $\mathbb{Z}_2$-twisted punctures

\begin{equation}
 \begin{matrix}\includegraphics[width=219pt]{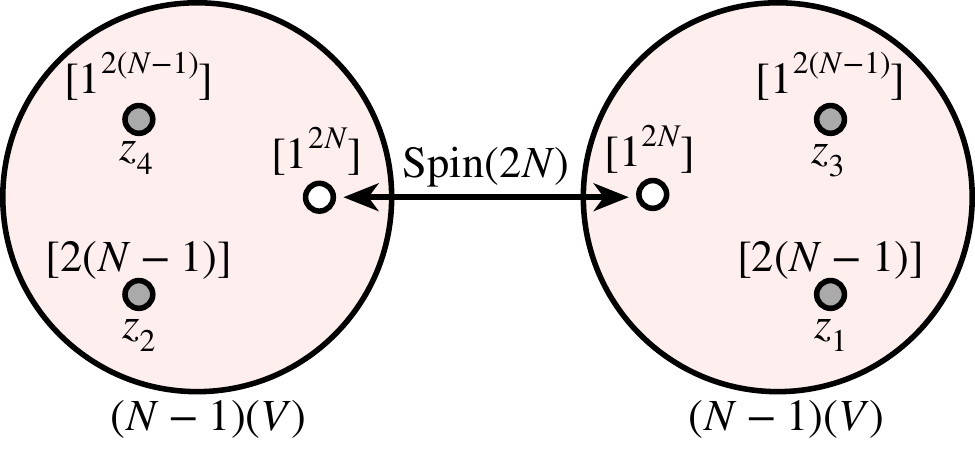}\end{matrix}\quad,
\label{SO2Nvectors}\end{equation}
there is a universal realization of $Spin(2N-1)$ with $2N-3$ fundamentals plus $(N-1)$ free hypermultiplets as a four-punctured sphere in the (twisted) $D_N$ theory

\begin{equation}
 \begin{matrix}\includegraphics[width=272pt]{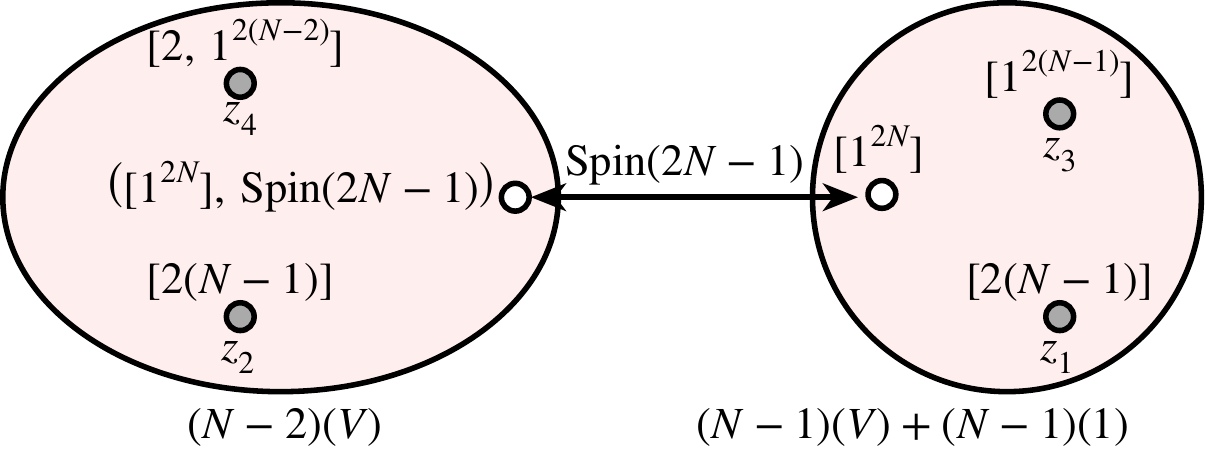}\end{matrix}\quad.
\label{SO2Nplus1vectors}\end{equation}
The Seiberg-Witten curve corresponding to \eqref{SO2Nvectors} takes the form of \eqref{SWDcurve} where the invariant $k$-differentials are

\begin{displaymath}
\begin{split}
  \phi_{2k}(z)&=  \frac{u_{2k}\, z_{1 4} z_{2 3} z_{3 4}^{2(k-1)} {(d z)}^{2k}}{(z-z_1)(z-z_2){(z-z_3)}^{2k-1}{(z-z_4)}^{2k-1}}\\
 \tilde{\phi}(z)&= \frac{\tilde{u}\, z_{1 4}^{1/2} z_{2 3}^{1/2} z_{3 4}^{N-1} {(d z)}^{N}}{{(z-z_1)}^{1/2}{(z-z_2)}^{1/2}{(z-z_3)}^{(2N-1)/2}{(z-z_4)}^{(2N-1)/2}}\quad.
\end{split}
\end{displaymath}
The Seiberg-Witten curve for \eqref{SO2Nplus1vectors} takes the same form, but with $\tilde{\phi}\equiv0$.

This pattern will repeat, in many of the examples below. The $Spin(2N-1)$ theory, with the same number of hypermultiplets in the spinor, but one fewer in the vector representation, is obtained by replacing the puncture at $z_4$, with one where the last box in the Young diagram is shifted to a new row. Physically, this corresponds to using one of the vector hypermultiplets to Higgs $Spin(2N)\to Spin(2N-1)$. The ``surprise'' is that integrating out the massive modes has such a simple effect on the Coulomb branch geometry.

The strong-coupling dual of \eqref{SO2Nplus1vectors} is an $SU(2)$ gauging of the $Sp(2N-3)_{2N-1}\times SU(2)_8$ SCFT, with $N-1$ additional free hypermultiplets

\begin{displaymath}
 \includegraphics[width=245pt]{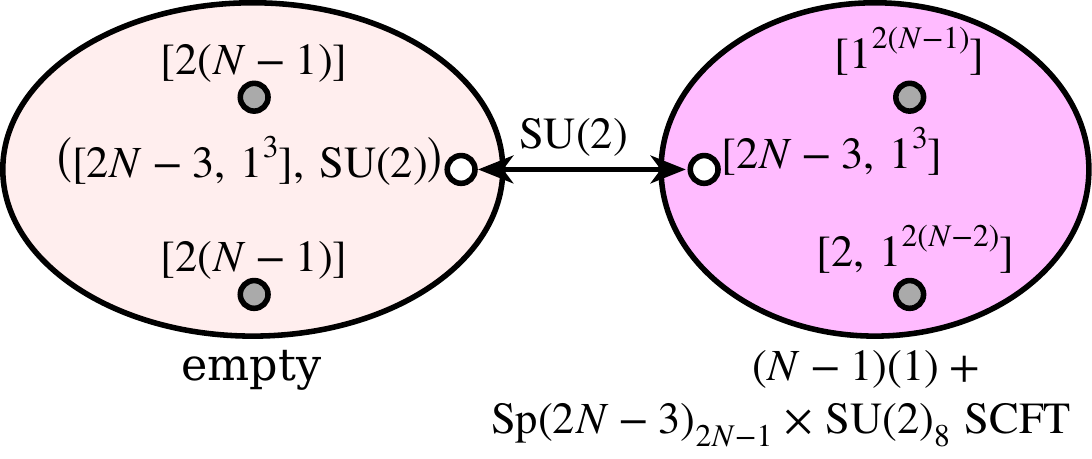}\quad.
\end{displaymath}
These theories have vanishing $\beta$-function for any $N$.

Including hypermultiplets in spinor representations will follow a similar pattern, where we will realize $Spin(2N-1)$ and $Spin(2N)$ gauge theories as 4-punctured spheres in the $D_N$ theory. The Seiberg-Witten curve for each of these theories takes the form \eqref{SWDcurve}. We list the invariant $k$-differentials for each theory below.

As we saw above, the solutions for $Spin(2N-1)$ is obtained from the corresponding $Spin(2N)$ theory (i.e, the theory with the same number of spinors (ignoring their chirality, for $N$ even) and one more vector) by setting $\tilde{u}=0$.

\section{$Spin(9)$ and $Spin(10)$ Gauge Theories}\label{_and__gauge_theories}
All of the following arise in the $D_5$ theory, possibly with $\mathbb{Z}_2$-twisted punctures.

\subsection{$Spin(9)$}\label{Spin9}
\subsubsection{$Spin(9)+1(16)+5(9)$}\label{Spin9_a}
\begin{equation}
 \begin{matrix}\includegraphics[width=277pt]{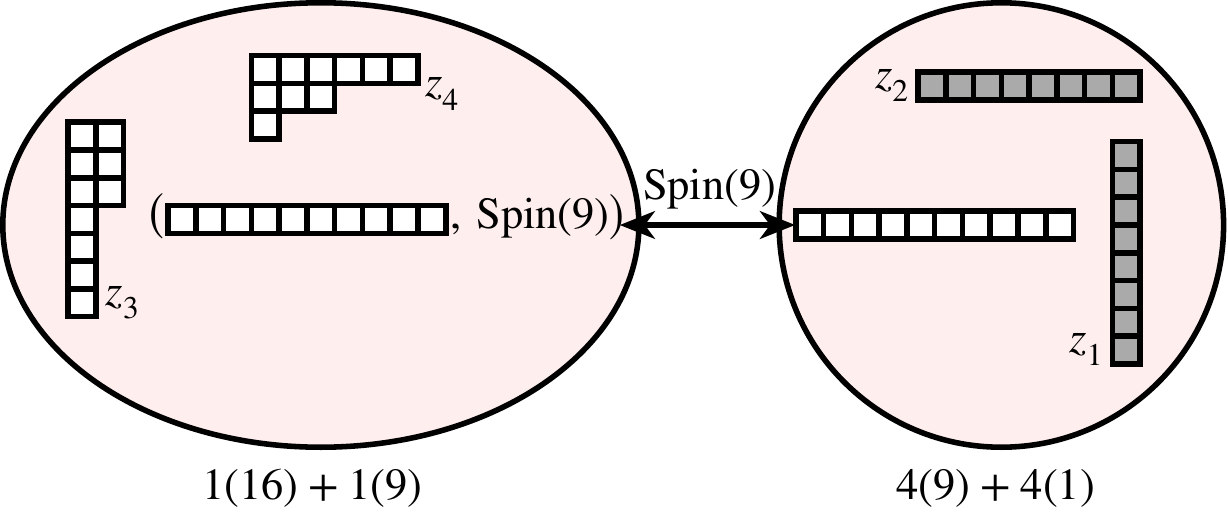}\end{matrix}\quad.
\label{SO95v1s}\end{equation}
The other degeneration limits yield a gauge theory fixture

\begin{displaymath}
 \includegraphics[width=254pt]{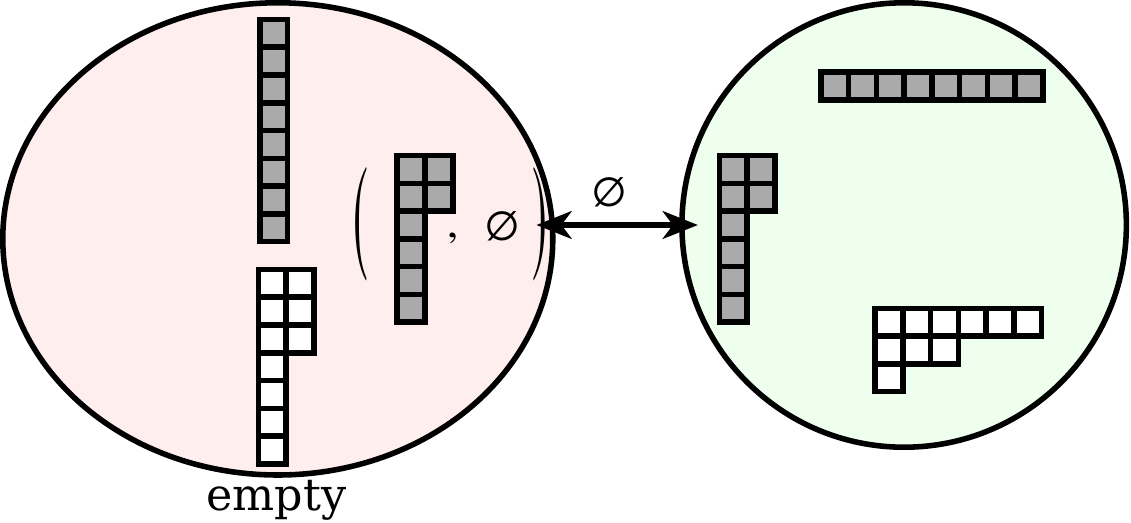}
\end{displaymath}
and an $Sp(2)$ gauging of the $Sp(7)_9\,\text{SCFT}+\tfrac{3}{2}(4)+4(1)$
\begin{displaymath}
 \begin{matrix}\includegraphics[width=254pt]{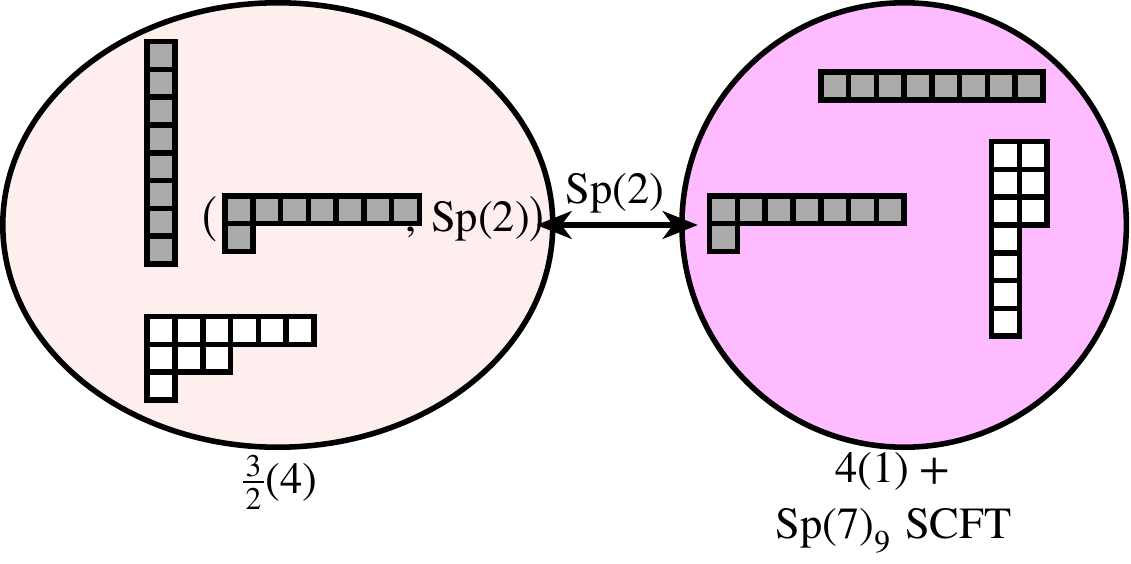}\end{matrix}\quad.
\end{displaymath}

\subsubsection{$Spin(9)+2(16)+3(9)$}\label{Spin9_b}
\begin{equation}
 \begin{matrix}\includegraphics[width=277pt]{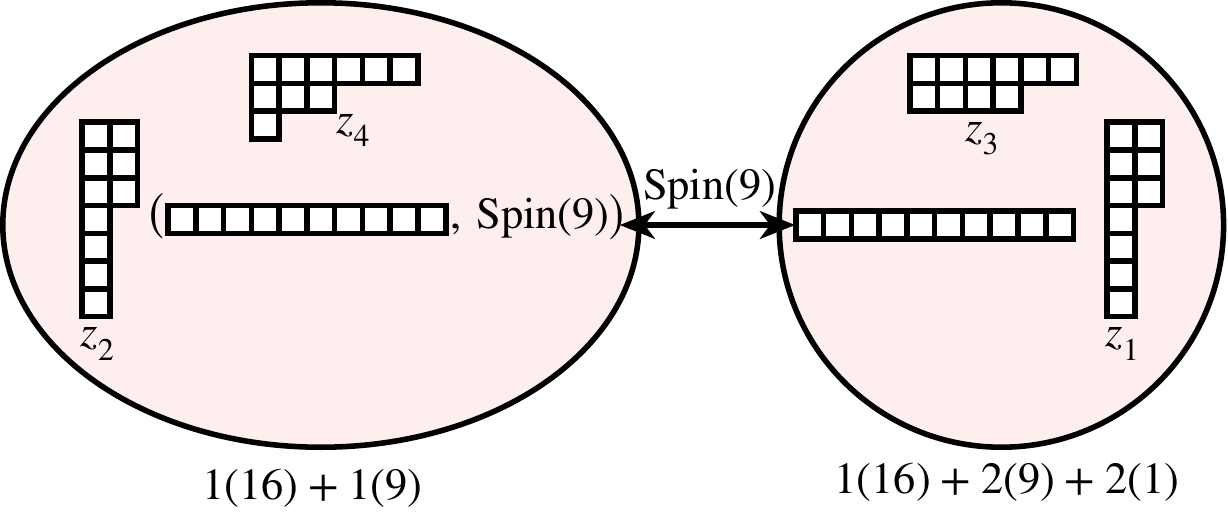}\end{matrix}\quad.
\label{SO93v2s}\end{equation}
The S-dual theory is an $SU(2)$ gauging of the $Sp(3)_{16} \times Sp(2)_9 \times SU(2)_7 \text{SCFT} + \frac{1}{2}(2)+2(1)$
\begin{displaymath}
 \begin{matrix}\includegraphics[width=331pt]{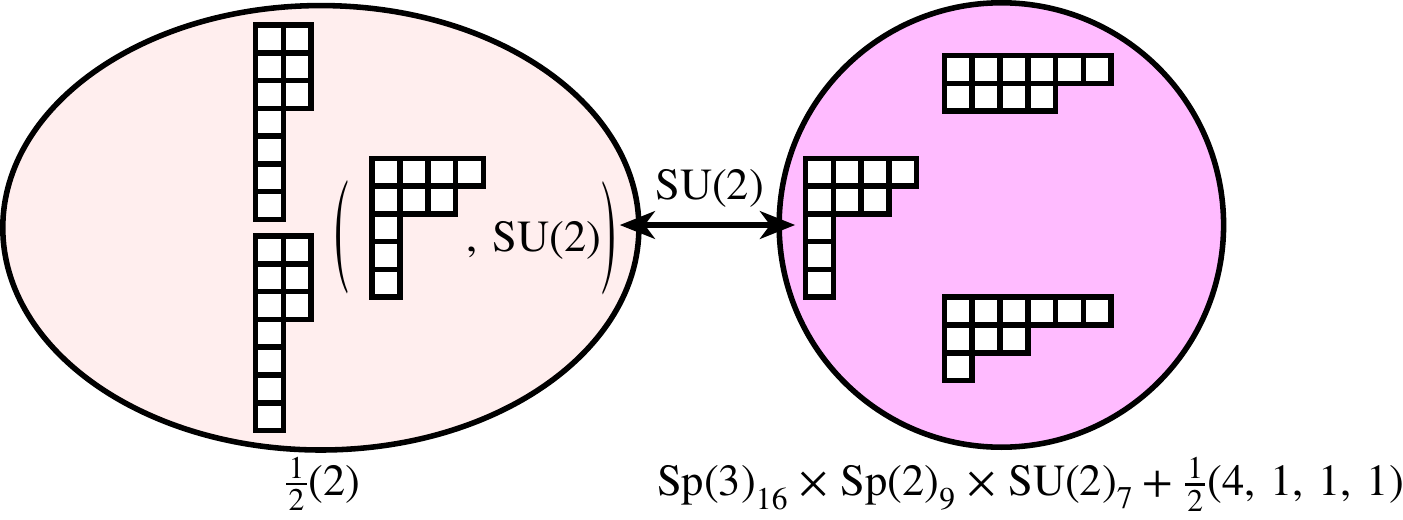}\end{matrix}.
\end{displaymath}

\subsubsection{$Spin(9)+3(16)+1(9)$}\label{Spin9_c}
\begin{equation}
 \begin{matrix}\includegraphics[width=277pt]{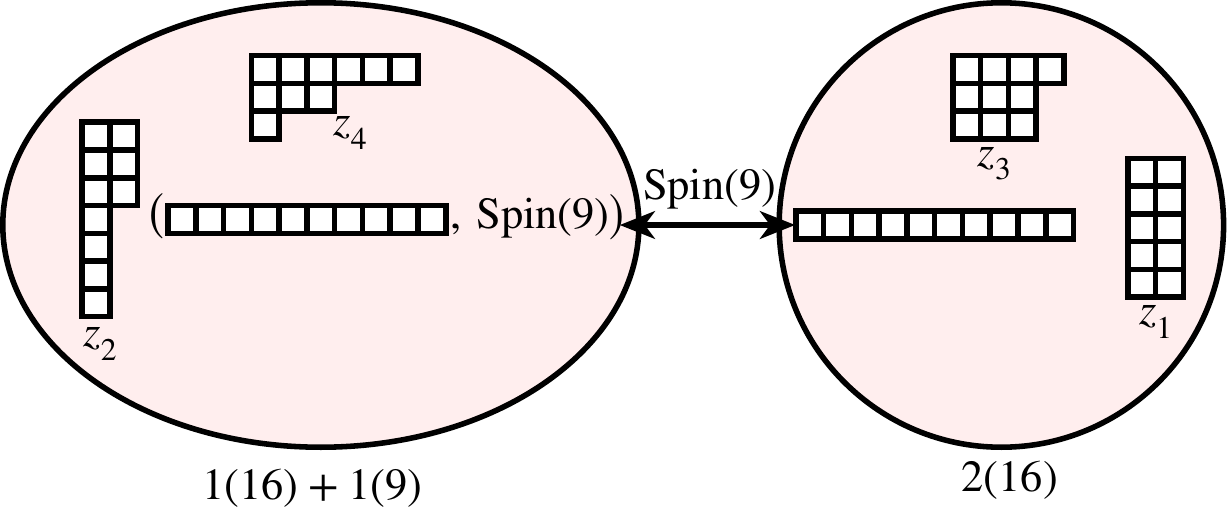}\end{matrix}\quad.
\label{SO91v3s}\end{equation}
The S-dual theories are an $SU(2)$ gauging of the $Sp(3)_{16} \times SU(2)_8 \times SU(2)_9$ SCFT
\begin{displaymath}
 \includegraphics[width=283pt]{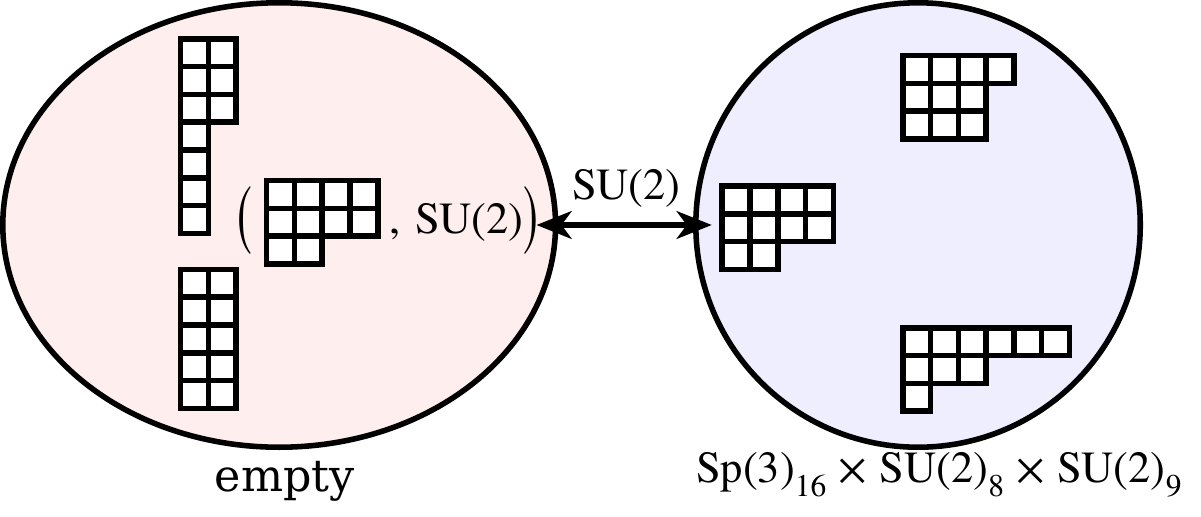}
\end{displaymath}
and a $G_2$ gauging of the ${(E_7)}_{16} \times SU(2)_9$ SCFT\footnote{This interacting fixture is another realization of the ${(E_7)}_{8n}\times {SU(2)}_{(n-1)(4n+1)}$ SCFT, which arises on the world volume of $n$ D3-branes probing a III${}^*$ singularity in F-theory (see \cite{Dasgupta:1996ij,Banks:1996nj,Aharony:2007dj} and \S5.3 of \cite{Chacaltana:2014jba}).}
\begin{displaymath}
 \begin{matrix}\includegraphics[width=258pt]{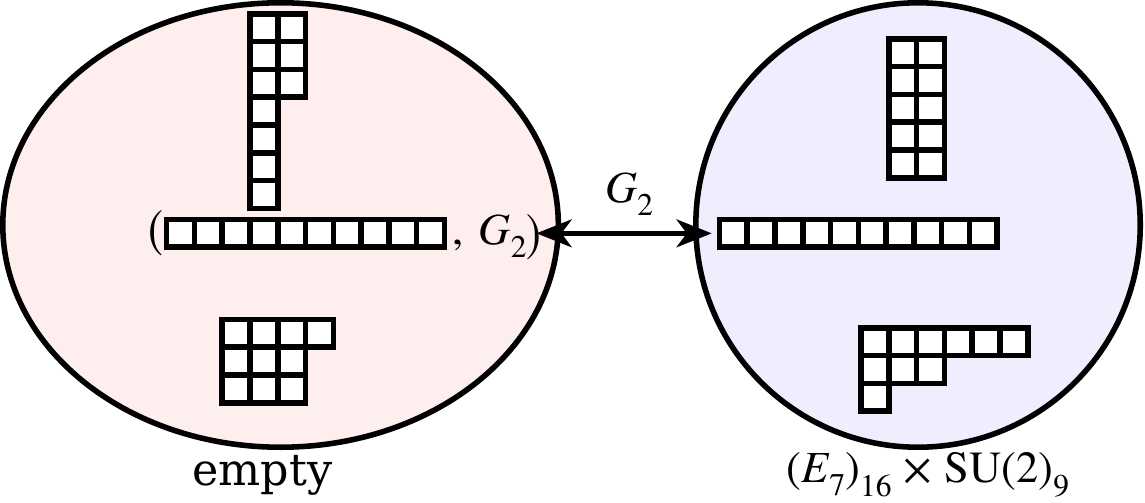}\end{matrix}\quad.
\end{displaymath}

\subsection{$Spin(10)$}\label{Spin10}
\subsubsection{$Spin(10)+1(16)+6(10)$}\label{Spin10_a}
\begin{equation}
 \begin{matrix}\includegraphics[width=239pt]{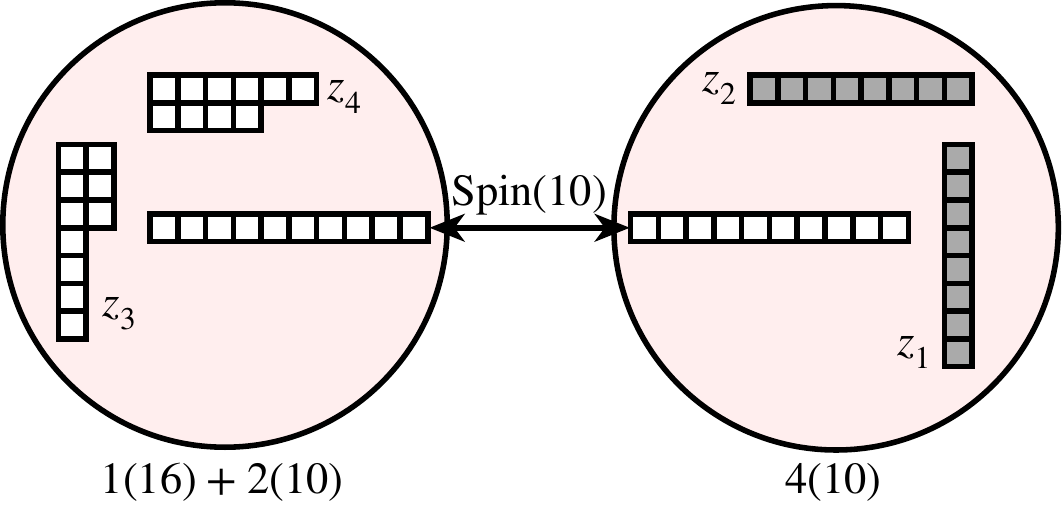}\end{matrix}\quad.
\label{SO106v1s}\end{equation}
The other degenerations involve a gauge theory fixture
\begin{displaymath}
 \includegraphics[width=239pt]{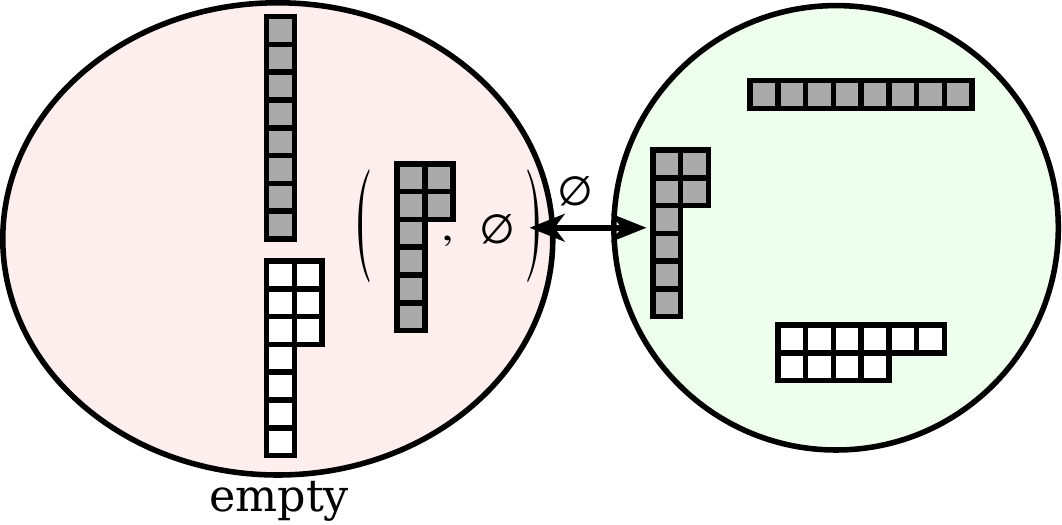}
\end{displaymath}
and an $Sp(2)$ gauging of the $Sp(8)_{10}\,\text{SCFT}+1(4)$
\begin{displaymath}
 \begin{matrix}\includegraphics[width=263pt]{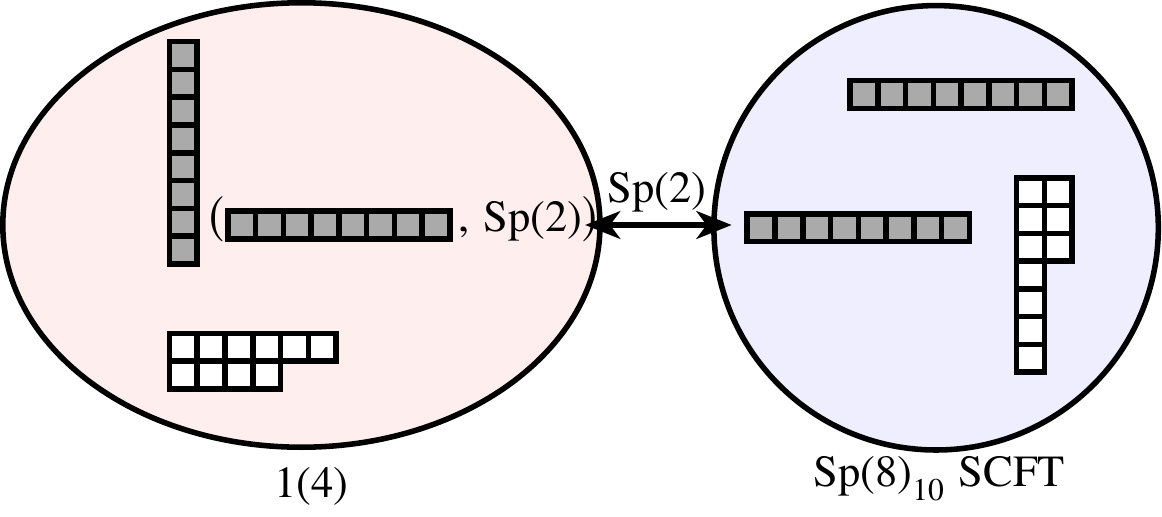}\end{matrix}\quad.
\end{displaymath}
The invariant $k$-differentials for \eqref{SO106v1s} are given by
\begin{align}\label{Spin10sol}
\phi_2(z)&=\frac{u_2\,z_{1 3}z_{2 4} {(d z)}^2}{(z-z_1)(z-z_2)(z-z_3)(z-z_4)}\notag\\
\phi_4(z)&=\frac{\left[u_4\, (z-z_3) z_{2 4} -\tfrac{1}{4}u_2^2\, (z-z_2) z_{3 4}\right]z_{1 3}z_{2 4}^2 {(d z)}^4}{(z-z_1){(z-z_2)}^3{(z-z_3)}^2{(z-z_4)}^3}\notag\\
\phi_6(z)&=\frac{u_6\,z_{1 3}z_{2 3}z_{2 4}^4 {(d z)}^6}{(z-z_1){(z-z_2)}^5{(z-z_3)}^2{(z-z_4)}^4}\\
\phi_8(z)&=\frac{u_8\,z_{1 3}z_{2 3}z_{2 4}^6 {(d z)}^8}{(z-z_1){(z-z_2)}^7{(z-z_3)}^2{(z-z_4)}^6}\notag\\
\tilde{\phi}(z)&= \frac{\tilde{u}\, z_{1 3}^{1/2} z_{2 3}^{1/2} z_{2 4}^{4} {(d z)}^5}{{(z-z_1)}^{1/2}{(z-z_2)}^{9/2}{(z-z_3)}{(z-z_4)}^{4}}\quad.\notag
\end{align}
The gauge theory moduli space is a branched double-cover of $\mathcal{M}_{0,4}$ and the gauge couplings are given by \eqref{twistedCoupling}.

The invariant $k$-differentials for \eqref{SO95v1s} are as above, but with $\tilde{\phi}\equiv 0$.

\subsubsection{$Spin(10)+2(16)+4(10)$}\label{Spin10_c}
\begin{equation}
 \begin{matrix}\includegraphics[width=239pt]{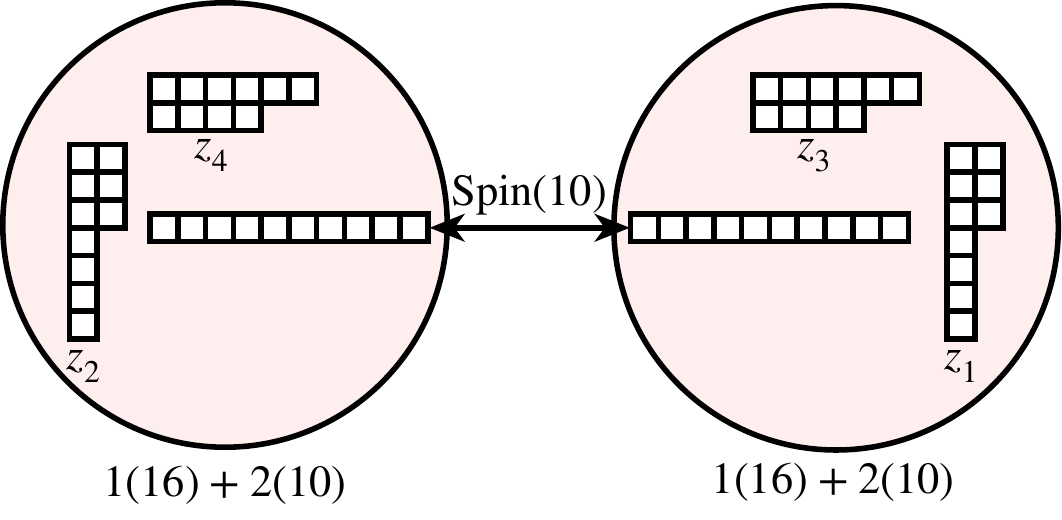}\end{matrix}
\label{SO104v2s}\end{equation}
The S-dual is an $SU(2)$ gauging of the $Sp(4)_{10} \times SU(2)_{16} \times SU(2)_7 \times U(1)\,\text{SCFT}+\tfrac{1}{2}(2)$
\begin{displaymath}
 \begin{matrix}\includegraphics[width=305pt]{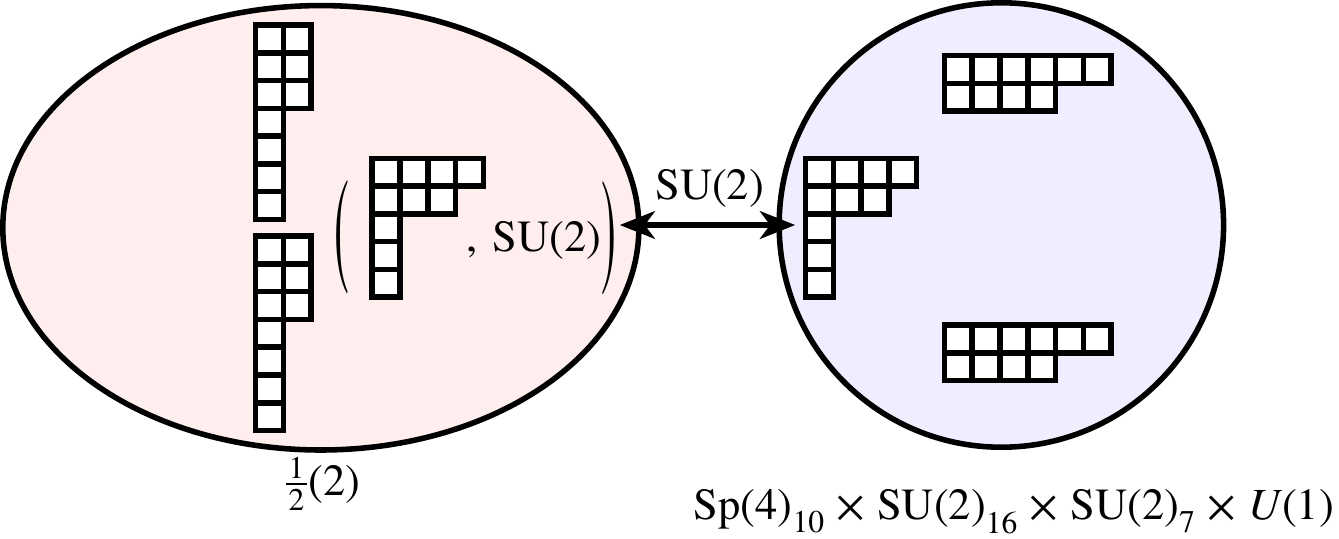}\end{matrix}.
\end{displaymath}
The invariant $k$-differentials for \eqref{SO104v2s} are given by
\begin{align}\label{SO104v2sSol}
\phi_2(z)&=\frac{u_2\, z_{1 2} z_{3 4} {(d z)}^2}{(z-z_1)(z-z_2)(z-z_3)(z-z_4)}\notag\\
\phi_4(z)&=\frac{\left[u_4\,(z-z_1)(z-z_2)z_{3 4} +\tfrac{1}{4} u_2^2\, \left((z-z_2)(z-z_3)z_{1 4}-(z-z_1)(z-z_4)z_{2 3}\right) \right]z_{1 2} z_{3 4}^2 {(d z)}^4}{{(z-z_1)}^2{(z-z_2)}^2{(z-z_3)}^3{(z-z_4)}^3}\notag\\
\phi_6(z)&=\frac{u_6\,z_{1 2}^2 z_{3 4}^4 {(d z)}^6}{{(z-z_1)}^2{(z-z_2)}^2{(z-z_3)}^4{(z-z_4)}^4}\\
\phi_8(z)&=\frac{u_8\, z_{1 2}^2 z_{3 4}^6 {(d z)}^8}{{(z-z_1)}^2{(z-z_2)}^2{(z-z_3)}^6{(z-z_4)}^6}\notag\\
\tilde{\phi}(z)&= \frac{\tilde{u}\, z_{1 2} z_{3 4}^4 {(d z)}^5}{{(z-z_1)}{(z-z_2)}{(z-z_3)}^4{(z-z_4)}^4}\quad.\notag
\end{align}
The $k$-differentials for \eqref{SO93v2s} are as above, but with $\tilde{\phi}\equiv 0$.

Since we are in the untwisted theory, the gauge theory moduli space is $\mathcal{M}_{0,4}$ (or more precisely, in this case, its $\mathbb{Z}_2$ quotient), and the gauge coupling is given by \eqref{untwistedCoupling}.

\subsubsection{$Spin(10)+3(16)+2(10)$}\label{Spin10_d}
\begin{equation}
 \begin{matrix}\includegraphics[width=239pt]{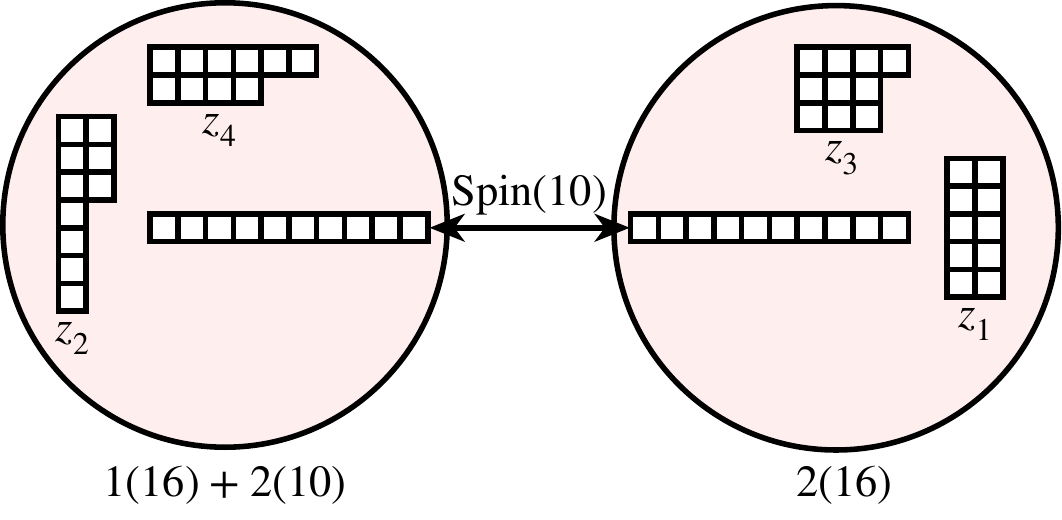}\end{matrix}\quad.
\label{SO102v3s}\end{equation}
The S-dual theories are an $SU(2)$ gauging of the $SU(3)_{32} \times Sp(2)_{10} \times SU(2)_8 \times U(1)$ SCFT
\begin{displaymath}
 \includegraphics[width=284pt]{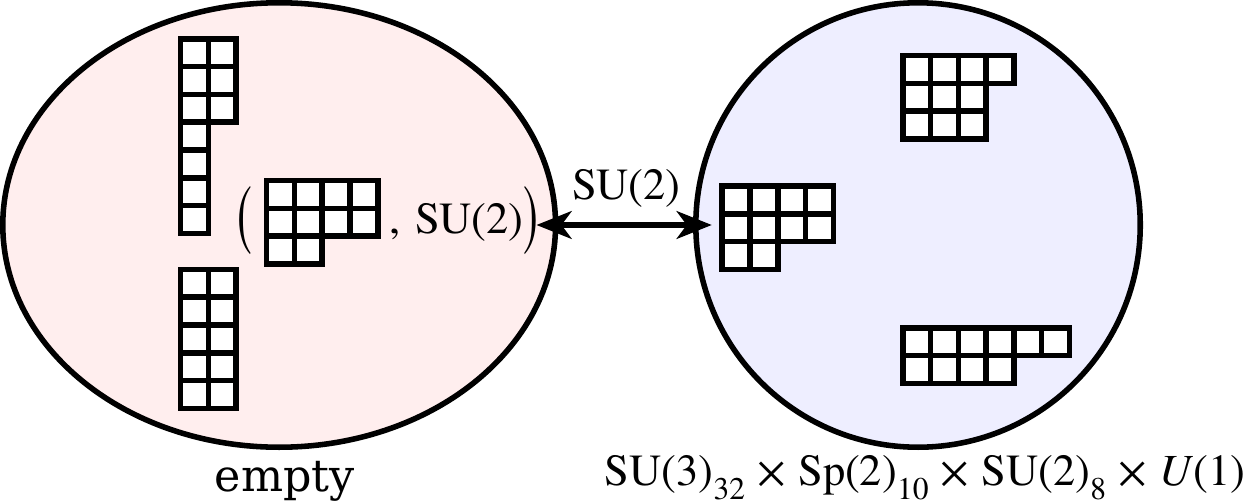}
\end{displaymath}
and a $G_2$ gauging of the $(E_6)_{16} \times Sp(2)_{10} \times U(1)$ SCFT
\begin{displaymath}
 \includegraphics[width=284pt]{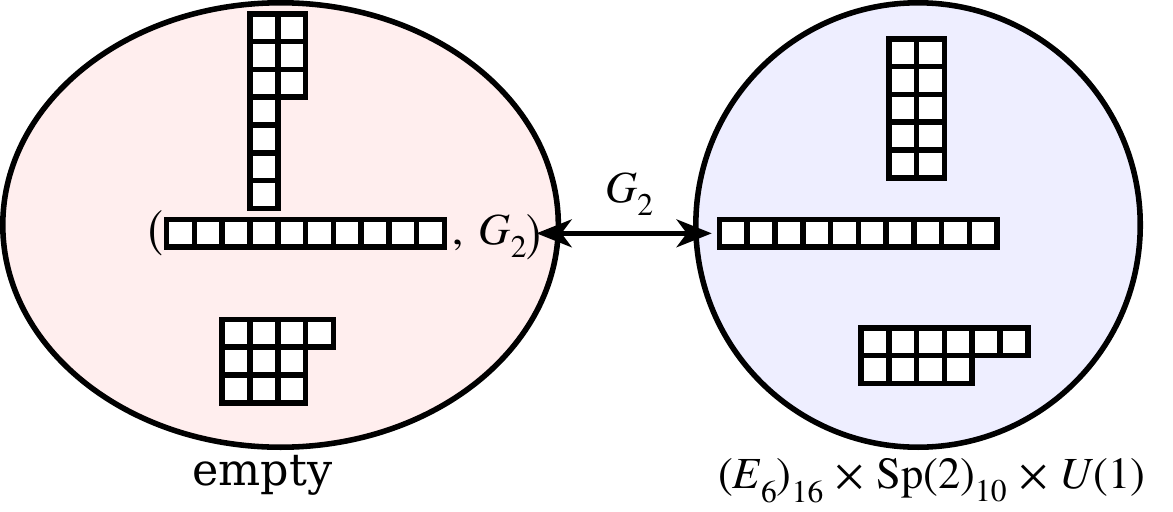}\quad.
\end{displaymath}
The invariant $k$-differentials for \eqref{SO102v3s} are given by
\begin{align}\label{SO102v3sSol}
\phi_2(z)&=\frac{u_2\, z_{1 2} z_{3 4} {(d z)}^2}{(z-z_1)(z-z_2)(z-z_3)(z-z_4)}\notag\\
\phi_4(z)&=\frac{\left[u_4\,(z-z_2)z_{1 4} +\tfrac{1}{4} u_2^2\, (z-z_4)z_{1 2} \right]z_{1 2} z_{3 4}^2 {(d z)}^4}{{(z-z_1)}^2{(z-z_2)}^2{(z-z_3)}^2{(z-z_4)}^3}\notag\\
\phi_6(z)&=\frac{\left[u_6\,(z-z_1)z_{3 4} +\tfrac{1}{2}u_2 u_4\, (z-z_4)z_{1 3} \right]z_{1 2}^2 z_{3 4}^3 {(d z)}^6}{{(z-z_1)}^3{(z-z_2)}^2{(z-z_3)}^4{(z-z_4)}^4}\\
\phi_8(z)&=\frac{\left[u_8\,(z-z_1)z_{3 4} +\tfrac{1}{4}u_4^2\, (z-z_4)z_{1 3} \right]z_{1 4} z_{1 2}^2 z_{3 4}^4 {(d z)}^8}{{(z-z_1)}^4{(z-z_2)}^2{(z-z_3)}^5{(z-z_4)}^6}\notag\\
\tilde{\phi}(z)&= \frac{\tilde{u}\, z_{1 4} z_{1 2} z_{3 4}^3 {(d z)}^5}{{(z-z_1)}^2{(z-z_2)}{(z-z_3)}^3{(z-z_4)}^4}\quad.\notag
\end{align}
The $k$-differentials for \eqref{SO91v3s} are as above, but with $\tilde{\phi}\equiv 0$.

\subsubsection{$Spin(10)+4(16)$}\label{Spin10_e}
\begin{equation}
 \begin{matrix}\includegraphics[width=239pt]{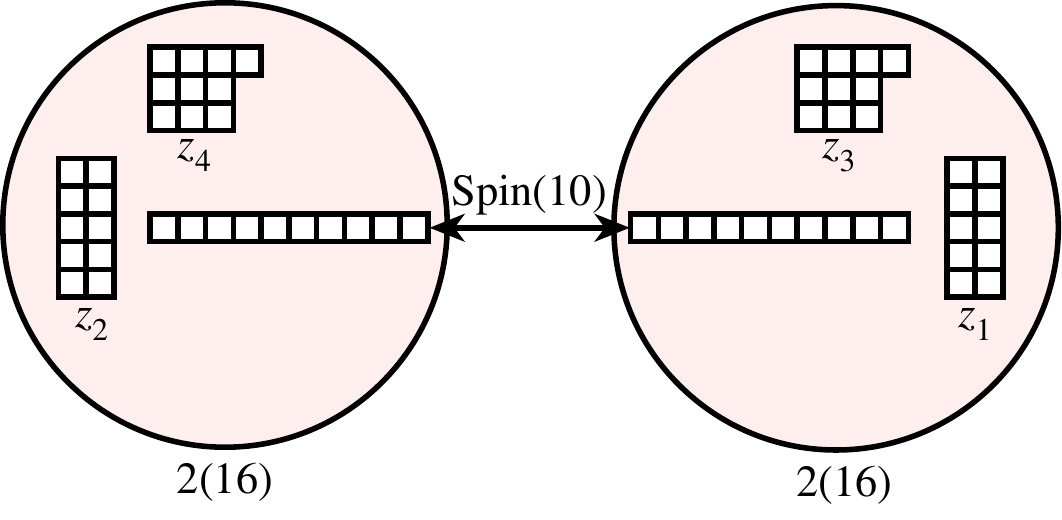}\end{matrix}\quad.
\label{SO104s}\end{equation}
The S-dual theory is an $Sp(2)$ gauging of the $SU(4)_{32} \times Sp(2)_{10}\,\text{SCFT}+1(4)$
\begin{displaymath}
 \begin{matrix}\includegraphics[width=239pt]{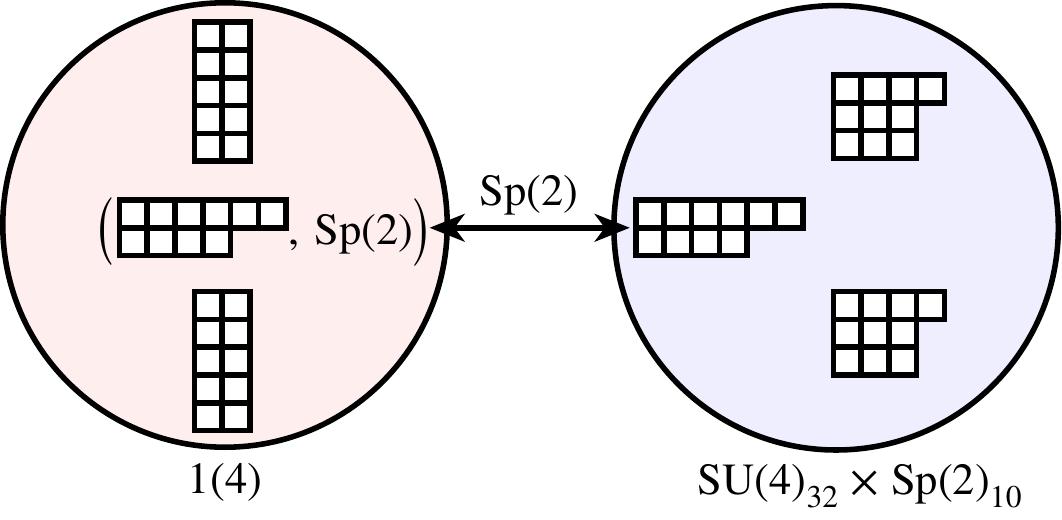}\end{matrix}\quad.
\end{displaymath}
For this theory, the $k$-differentials characterizing the Seiberg-Witten curve are
\begin{align}\label{Spin10sol2}
\phi_2(z)&=\frac{u_2\, z_{1 2} z_{3 4} {(d z)}^2}{(z-z_1)(z-z_2)(z-z_3)(z-z_4)}\notag\\
\phi_4(z)&=\frac{u_4\, z_{1 2}^2 z_{3 4}^2 {(d z)}^4}{{(z-z_1)}^2{(z-z_2)}^2{(z-z_3)}^2{(z-z_4)}^2}\notag\\
\phi_6(z)&=\frac{\left[u_6\,(z-z_1)(z-z_2)z_{3 4} -\tfrac{1}{2}u_2 \left(u_4-\tfrac{1}{4} u_2^2\right)\left((z-z_1)(z-z_3)z_{2 4}-(z-z_2)(z-z_4)z_{1 3}\right) \right]z_{1 2}^2 z_{3 4}^3 {(d z)}^6}{{(z-z_1)}^3{(z-z_2)}^3{(z-z_3)}^4{(z-z_4)}^4}\notag\\
\phi_8(z)&=\frac{\left[u_8\,(z-z_1)(z-z_2)z_{3 4} -\tfrac{1}{4}{\left(u_4-\tfrac{1}{4}u_2^2\right)}^2\left((z-z_1)(z-z_3)z_{2 4}-(z-z_2)(z-z_4)z_{1 3}\right)\right]z_{1 2}^3 z_{3 4}^4 {(d z)}^8}{{(z-z_1)}^4{(z-z_2)}^4{(z-z_3)}^5{(z-z_4)}^5}\notag\\
\tilde{\phi}(z)&= \frac{\tilde{u}\,  z_{1 2}^2 z_{3 4}^3 {(d z)}^5}{{(z-z_1)}^2{(z-z_2)}^2{(z-z_3)}^3{(z-z_4)}^3}\quad.
\end{align}
In this case, there are no hypermultiplets in the vector, which one could use to Higgs $Spin(10)\to Spin(9)$. Equivalently, it's not possible to move the last box, in the Young diagram at $z_4$, to a new row while keeping it a D-partition. So there is no corresponding $Spin(9)$ gauge theory.

\section{$Spin(11)$ and $Spin(12)$ Gauge Theories}\label{Spin11_and_Spin12}
These arise in the compactification of the $D_6$ theory, possibly with $\mathbb{Z}_2$-twisted punctures.

\subsection{$Spin(11)$}\label{Spin11}
\subsubsection{$Spin(11)+\tfrac{1}{2}(32)+7(11)$}\label{Spin11_a}
\begin{equation}
  \begin{matrix}\includegraphics[width=297pt]{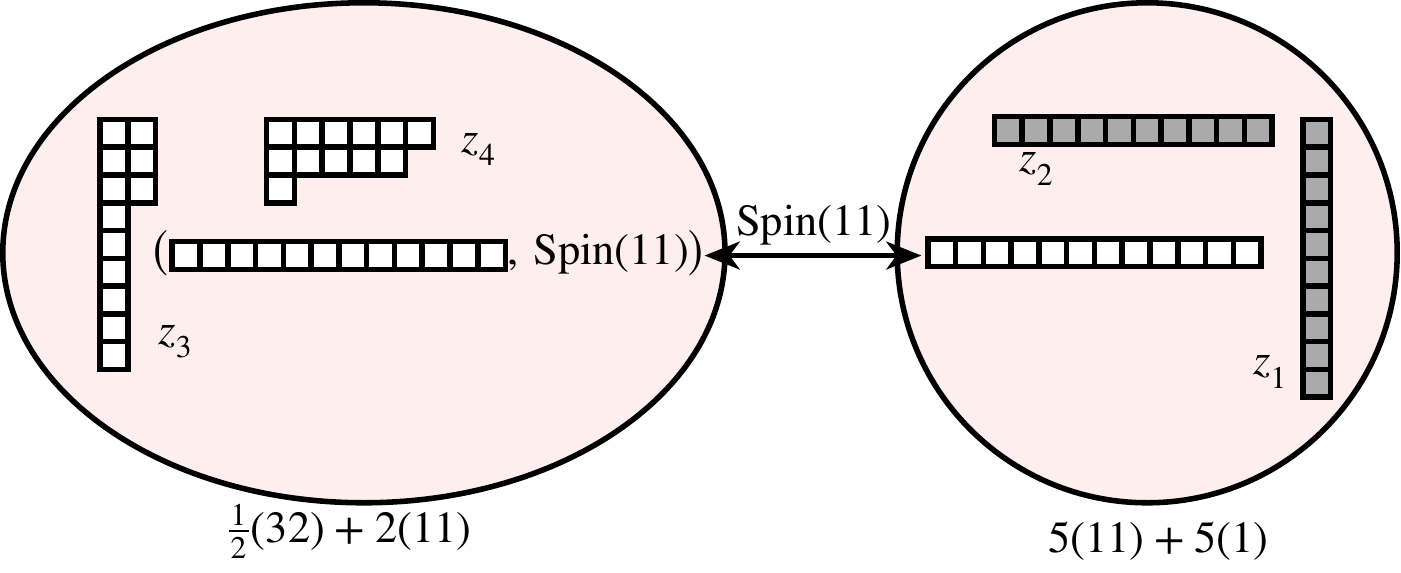}\end{matrix}\quad.
\label{SO117v1s}\end{equation}
The other degenerations involve a gauge theory fixture
\begin{displaymath}
 \begin{matrix}\includegraphics[width=283pt]{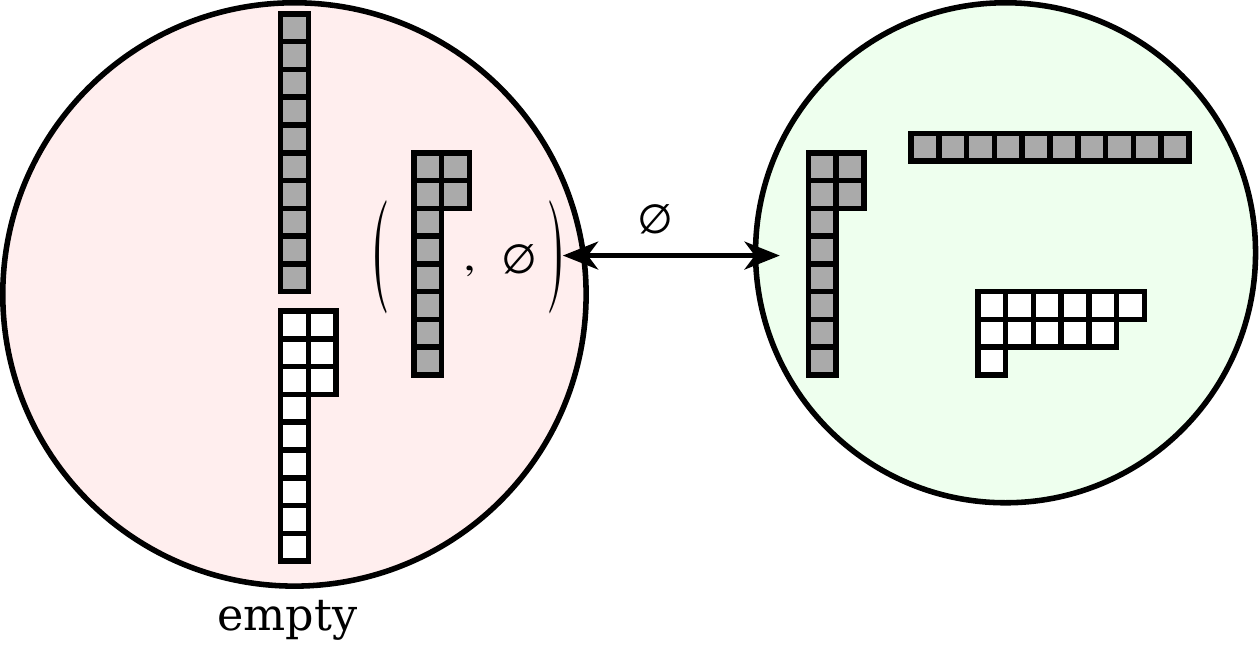}\end{matrix}
\end{displaymath}
and an $Sp(2)$ gauging of the ${Sp(9)}_{11}\,\text{SCFT}\,+\tfrac{1}{2}(4)+5(1)$
\begin{displaymath}
 \begin{matrix}\includegraphics[width=297pt]{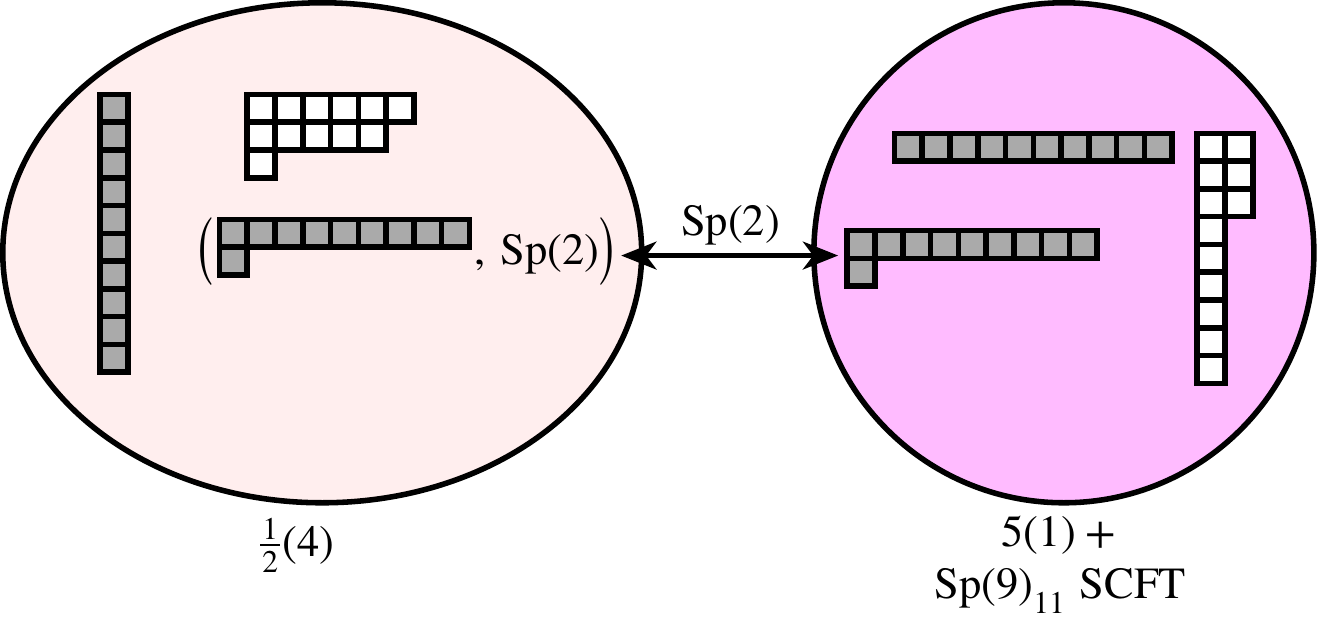}\end{matrix}\quad.
\end{displaymath}

\subsubsection{$Spin(11)+1(32)+5(11)$}\label{Spin11_a}
\begin{equation}
 \begin{matrix}\includegraphics[width=315pt]{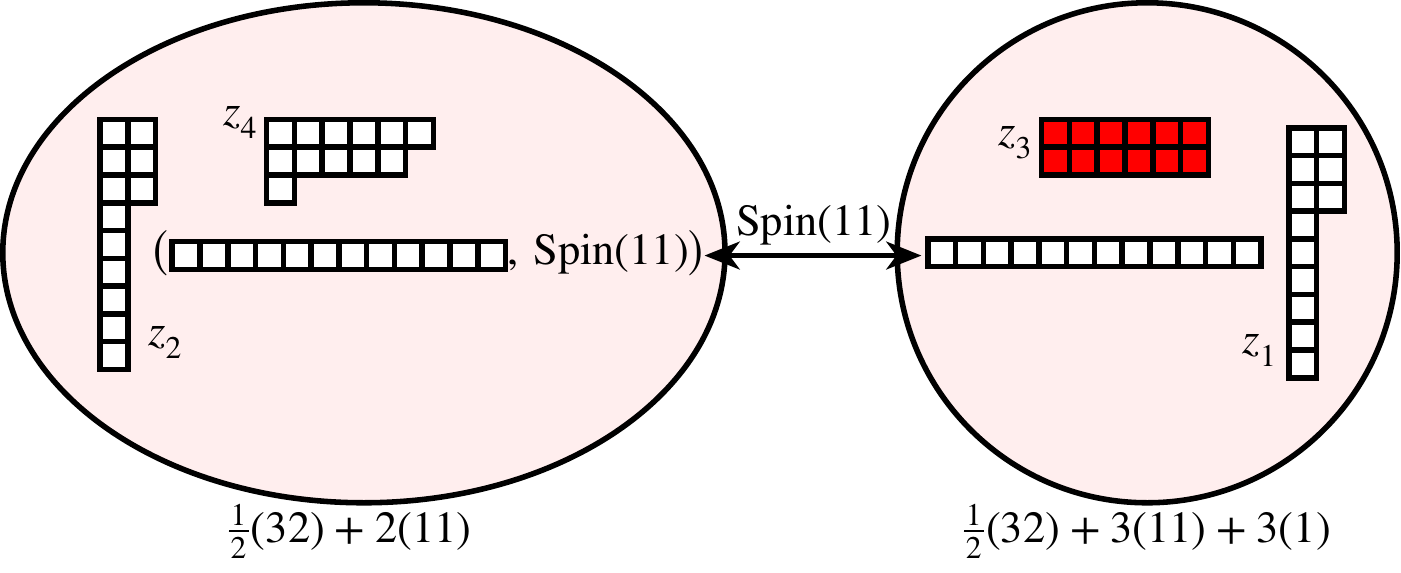}\end{matrix}\quad.
\label{SO115v2s}\end{equation}
The S-dual theory is an $SU(2)$ gauging of the $Sp(5)_{11} \times SU(2)_7 \times U(1)\, \text{SCFT}\, + \frac{1}{2}(2) +3(1)$,
\begin{displaymath}
 \begin{matrix}\includegraphics[width=290pt]{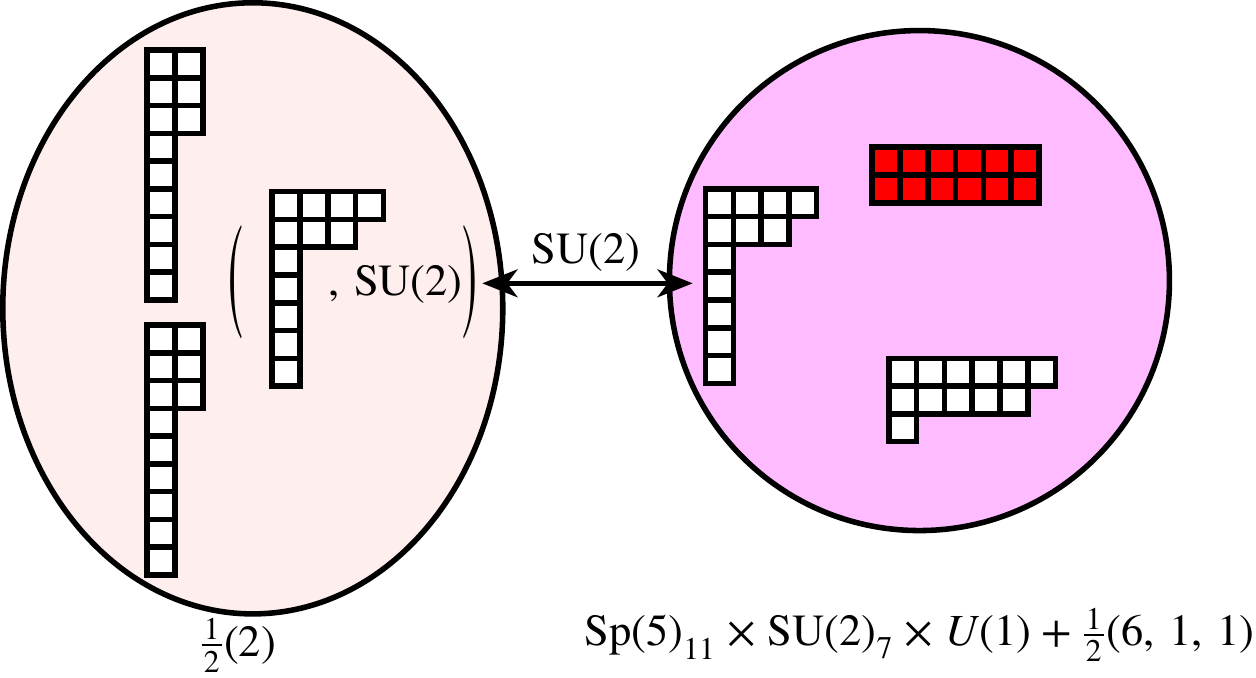}\end{matrix}.
\end{displaymath}

\subsubsection{$Spin(11)+\tfrac{3}{2}(32)+3(11)$}\label{Spin11_b}
\begin{equation}
 \begin{matrix}\includegraphics[width=315pt]{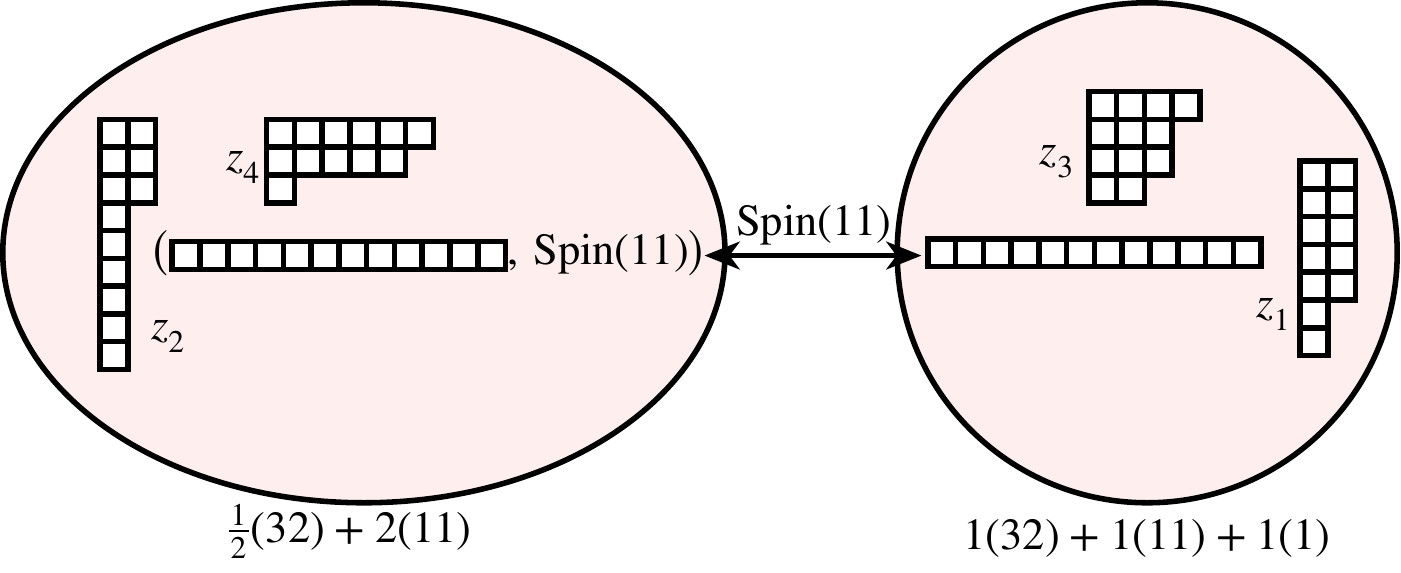}\end{matrix}\quad.
\label{SO113v3s}\end{equation}
The S-dual theories are an $SU(2)$ gauging of the $Sp(3)_{11} \times SU(2)_8 \times SU(2)_{128}\, \text{SCFT}\, + 1(1)$

\begin{displaymath}
 \includegraphics[width=367pt]{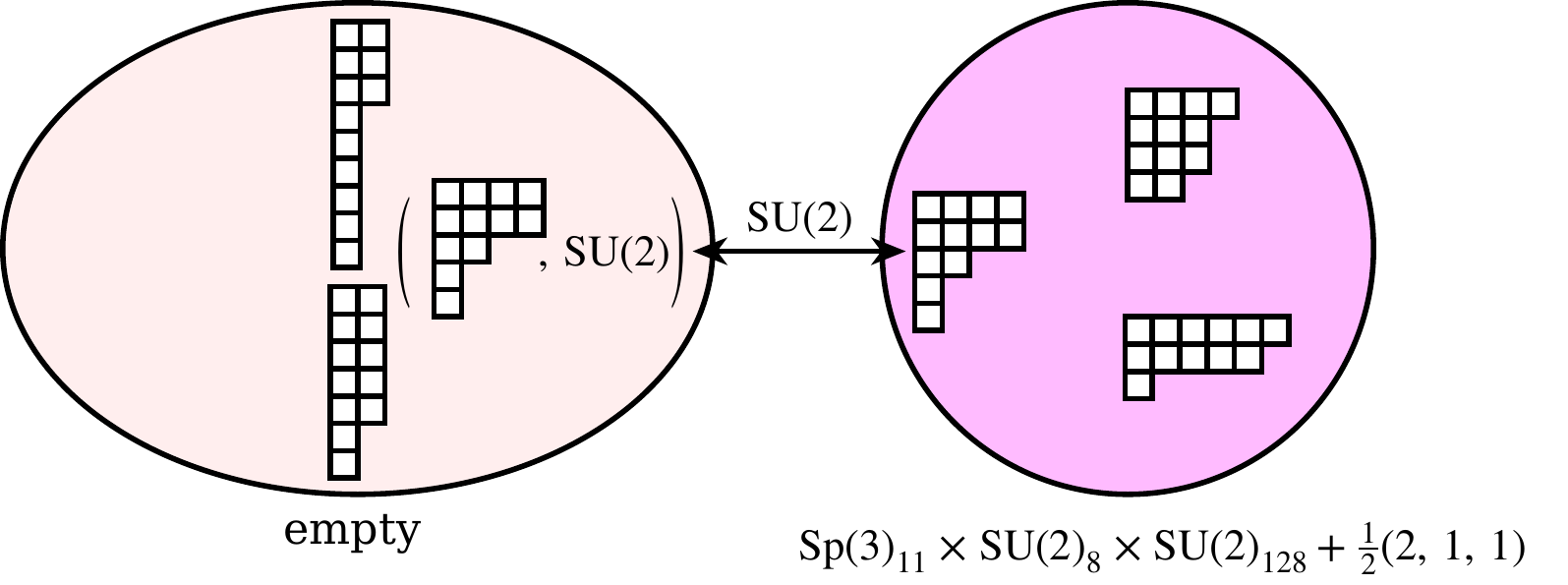}
\end{displaymath}
and a $G_2$ gauging\footnote{Note that, here, we use the Lie algebra embedding, $(\mathfrak{f}_4)_k\supset (\mathfrak{g}_2)_k\times \mathfrak{su}(2)_{8k}$.} of the $Sp(3)_{11} \times (F_4)_{16}\, \text{SCFT}\, + 1(1)$
\begin{displaymath}
 \begin{matrix}\includegraphics[width=347pt]{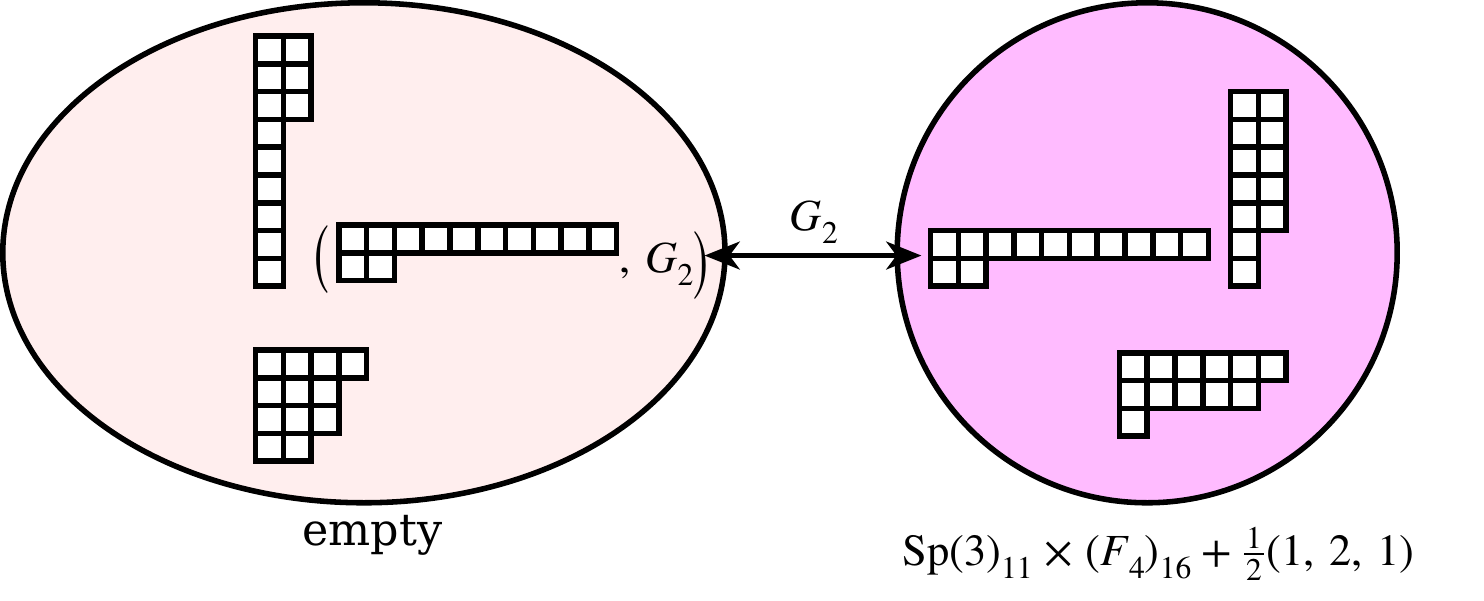}\end{matrix}.
\end{displaymath}

\subsubsection{$Spin(11)+2(32)+1(11)$}\label{Spin11_c}
\begin{equation}
 \begin{matrix}\includegraphics[width=315pt]{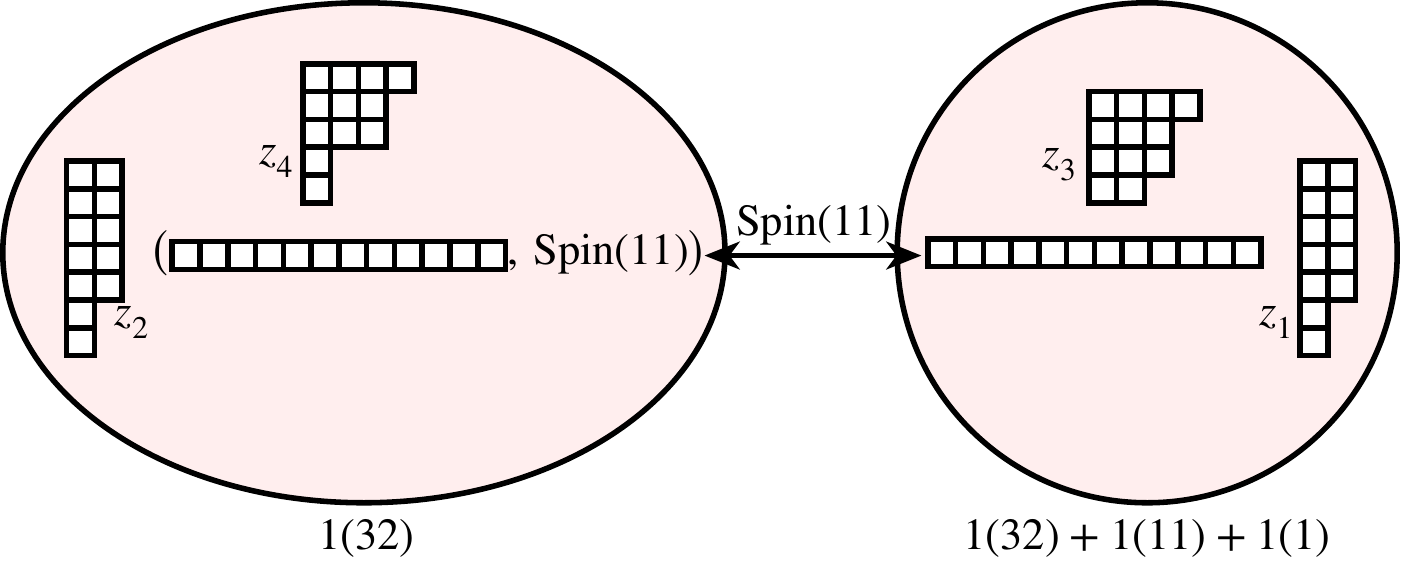}\end{matrix}\quad.
\label{SO111v4s}\end{equation}
The S-dual theory is an $Sp(2)$ gauging of the $Sp(3)_{11} \times {SU(2)}^2_{32}\, \text{SCFT}+\frac{1}{2}(4)+1(1)$

\begin{displaymath}
 \begin{matrix}\includegraphics[width=292pt]{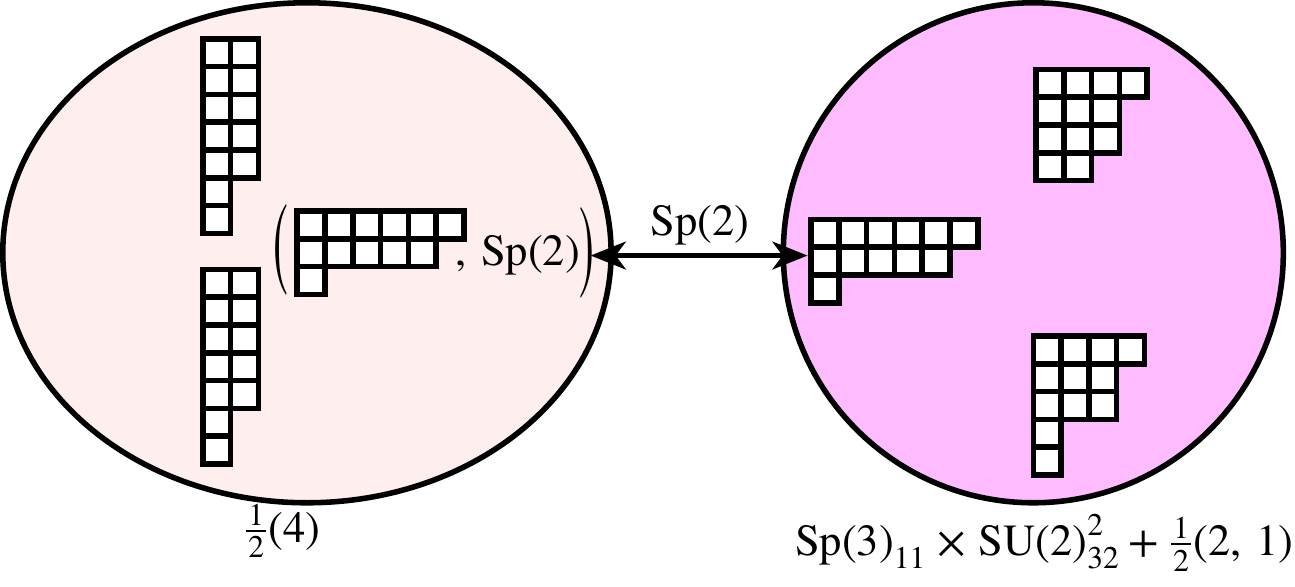}\end{matrix}\quad.
\end{displaymath}

\subsection{$Spin(12)$}\label{Spin12}
\subsubsection{$Spin(12)+\tfrac{1}{2}(32)+8(12)$}\label{Spin12_a}
\begin{equation}
 \begin{matrix}\includegraphics[width=264pt]{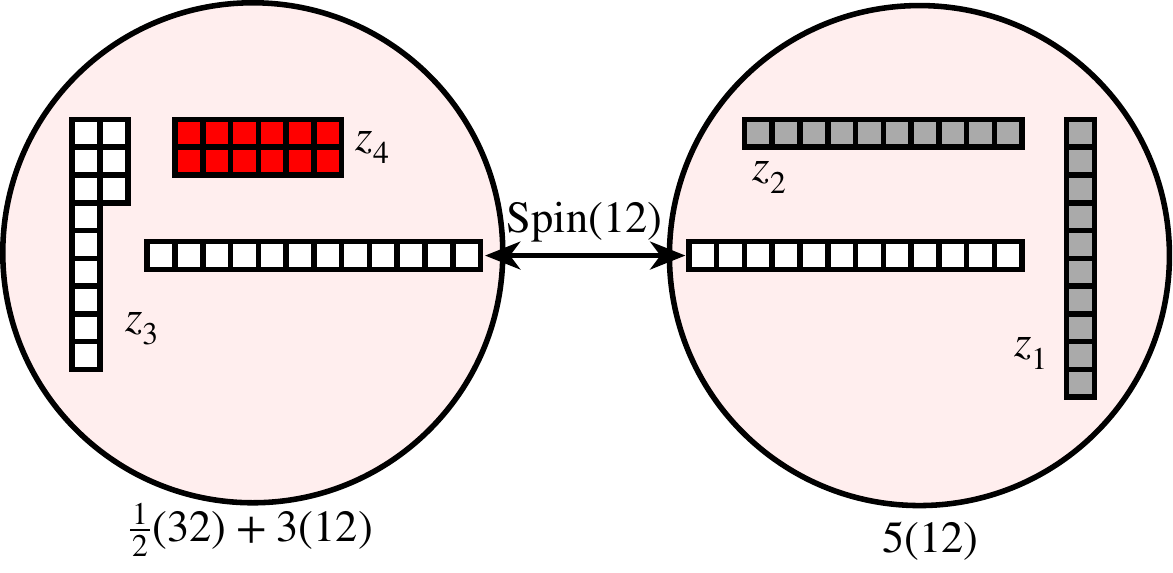}\end{matrix}\quad.
\label{SO128v1s}\end{equation}
The other degenerations involve an gauge theory fixture
\begin{displaymath}
 \includegraphics[width=264pt]{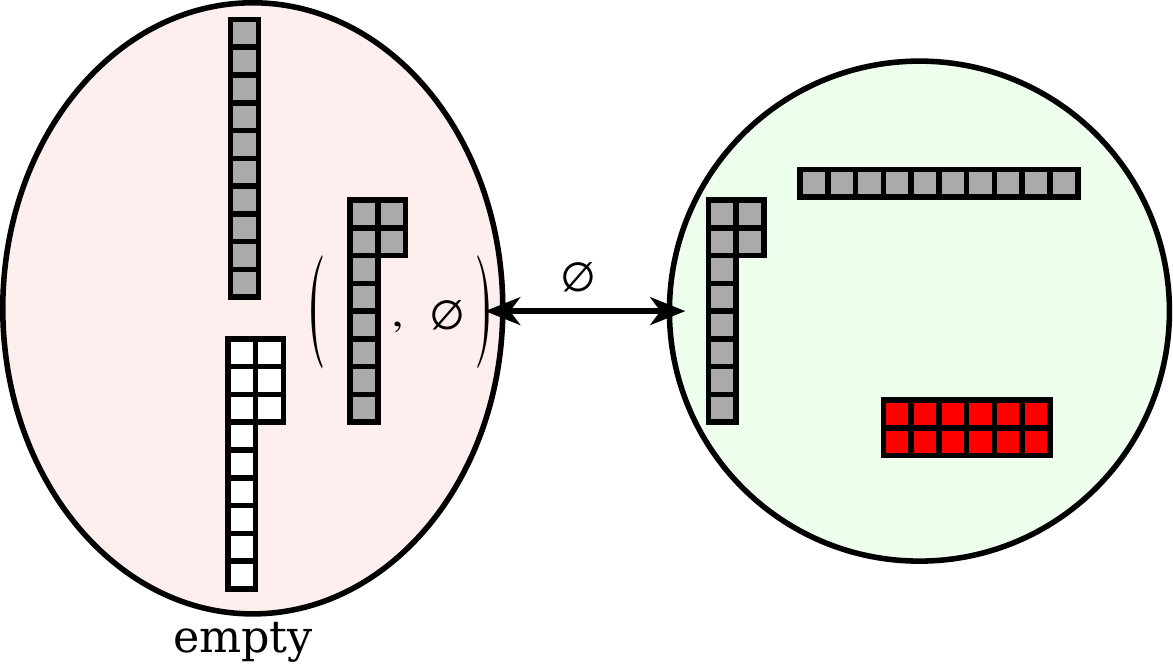}
\end{displaymath}
and an $Sp(2)$ gauging of the $Sp(10)_{12}$ SCFT

\begin{displaymath}
 \begin{matrix}\includegraphics[width=289pt]{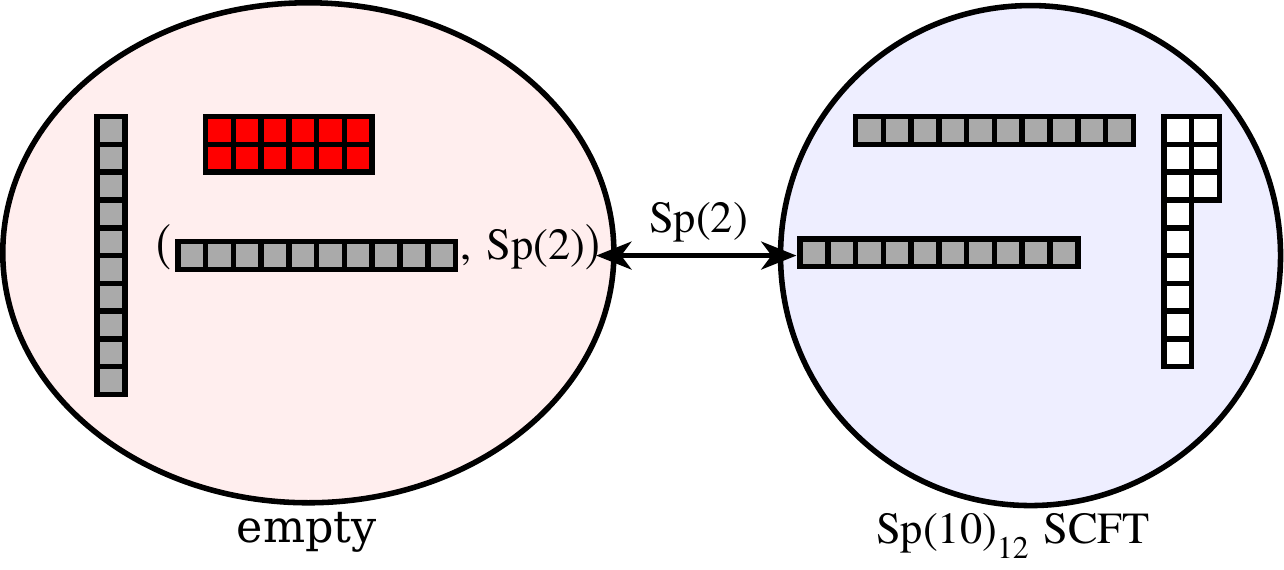}\end{matrix}\quad.
\end{displaymath}
The invariant $k$-differentials for \eqref{SO128v1s} are given by
\begin{align}\label{SO128v1sSol}
\phi_2(z)&=\frac{u_2\,z_{1 3}z_{2 4} {(d z)}^2}{(z-z_1)(z-z_2)(z-z_3)(z-z_4)}\notag\\
\phi_4(z)&=\frac{\left[u_4\, (z-z_3) z_{2 4} -\tfrac{1}{4}u_2^2\, (z-z_2) z_{3 4}\right]z_{1 3}z_{2 4}^2 {(d z)}^4}{(z-z_1){(z-z_2)}^3{(z-z_3)}^2{(z-z_4)}^3}\notag\\
\phi_6(z)&=\frac{[u_6\, (z-z_4)z_{13}+2\tilde{u}(z-z_3)z_{14}]z_{23}{z_{24}}^4{(d z)}^6}{{(z-z_1)}{(z-z_2)}^5{(z-z_3)}^2{(z-z_4)}^5}\notag\\
\phi_8(z)&=\frac{u_8\,z_{1 3}z_{2 3}z_{2 4}^6 {(d z)}^8}{(z-z_1){(z-z_2)}^7{(z-z_3)}^2{(z-z_4)}^6}\\
\phi_{10}(z)&=\frac{u_{10}\,  z_{13}z_{23}{z_{24}}^8{(d z)}^{10}}{(z-z_1){(z-z_2)}^9{(z-z_3)}^2{(z-z_4)}^8}\notag\\
\tilde{\phi}(z)&=\frac{\tilde{u}{z_{14}}^{1/2}{z_{24}}^{9/2}z_{23}{(d z)}^6}{{(z-z_1)}^{1/2}{(z-z_2)}^{11/2}{(z-z_3)}{(z-z_4)}^5}\quad.\notag
\end{align}
For \eqref{SO117v1s}, they are as above, but with $\tilde{u}\equiv 0$.

\subsubsection{$Spin(12)+1(32)+6(12)$}\label{Spin12_b}
\begin{equation}
 \begin{matrix}\includegraphics[width=264pt]{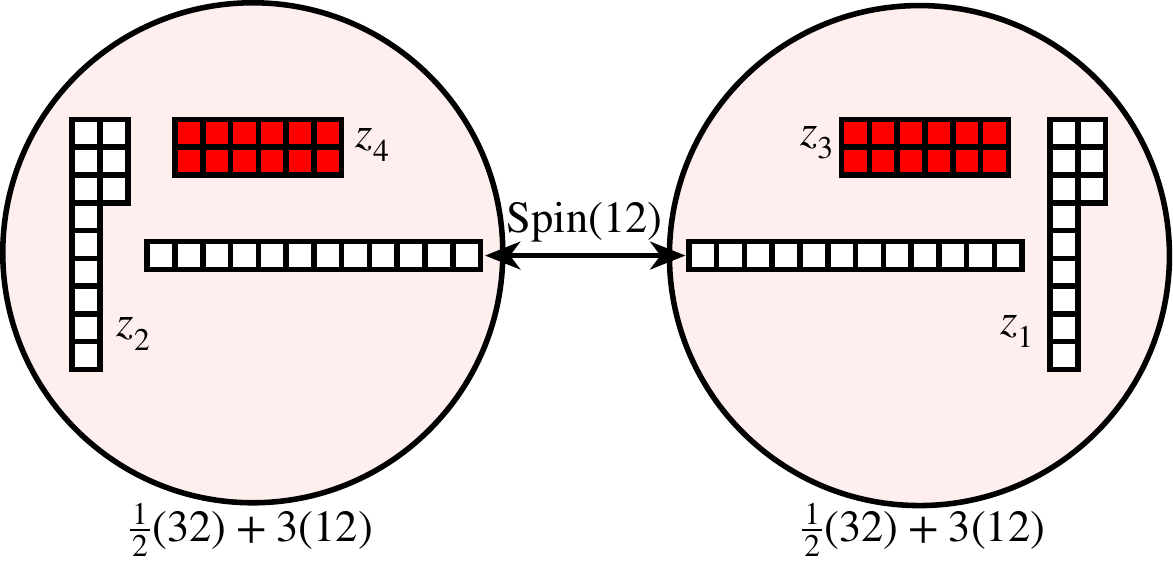}\end{matrix}\quad.
\label{SO126v2s}\end{equation}
The S-dual theory is an $SU(2)$ gauging of the $Sp(6)_{12} \times SU(2)_7 \times U(1)\,\text{SCFT}\,+\tfrac{1}{2}(2)$

\begin{displaymath}
 \begin{matrix}\includegraphics[width=264pt]{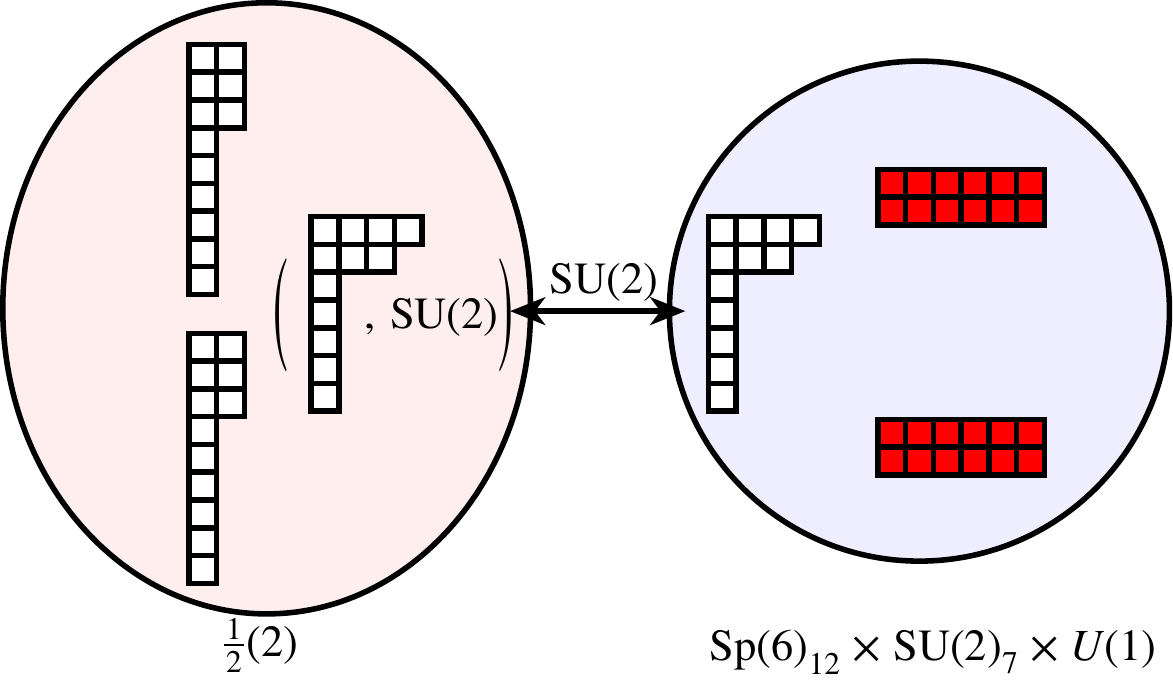}\end{matrix}\quad.
\end{displaymath}

\subsubsection{$Spin(12)+\tfrac{1}{2}(32)+\tfrac{1}{2}(32')+6(12)$}\label{Spin12_b}
\begin{equation}
 \begin{matrix}\includegraphics[width=264pt]{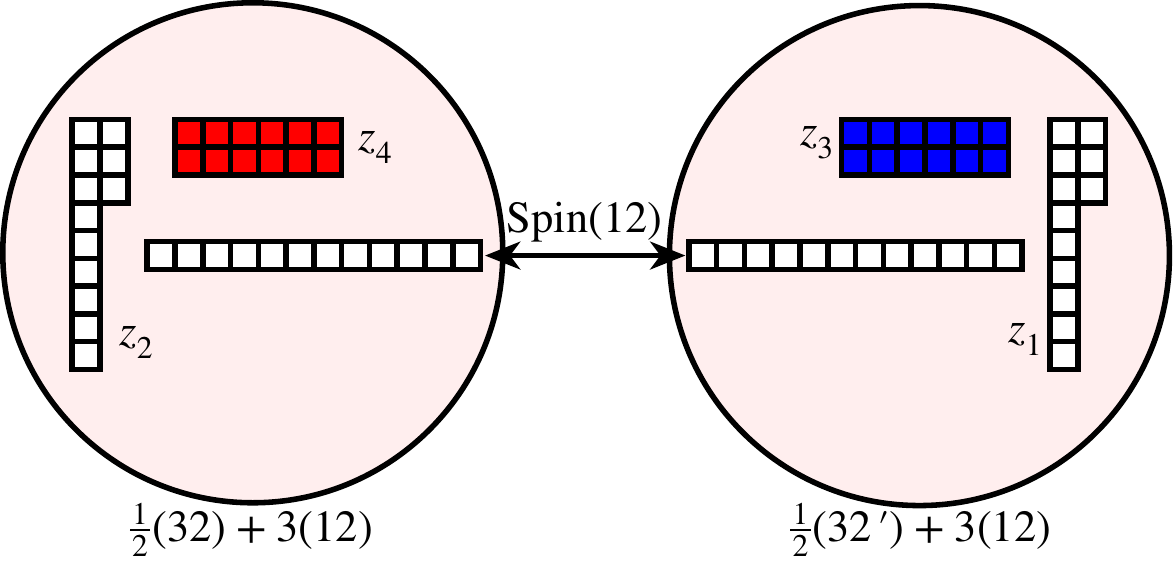}\end{matrix}\quad.
\label{SO126v1s1sp}\end{equation}
The S-dual is an $SU(2)$ gauging of the $Sp(6)_{12} \times SU(2)_7\,\text{SCFT}\,+\tfrac{1}{2}(2)$
\begin{displaymath}
 \begin{matrix}\includegraphics[width=264pt]{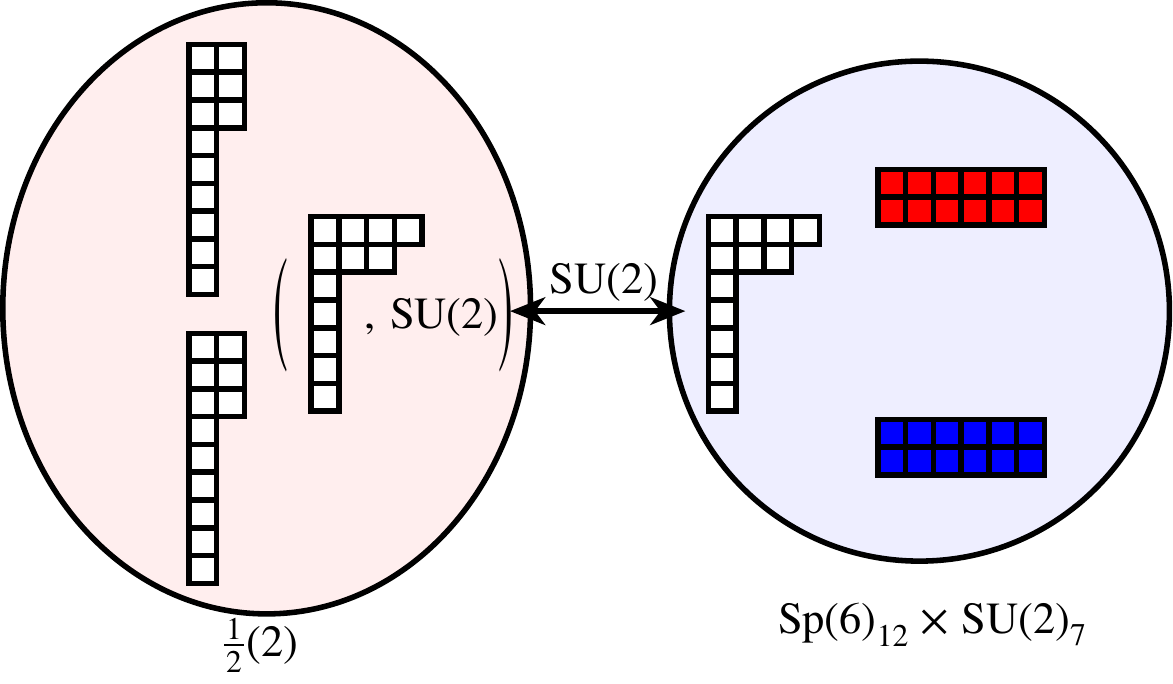}\end{matrix}\quad.
\end{displaymath}
The invariant $k$-differentials for \eqref{SO126v2s} and \eqref{SO126v1s1sp} are
\begin{align}\label{SO126v2sSol}
\phi_2(z)&=\frac{u_2\, z_{1 2} z_{3 4} {(d z)}^2}{(z-z_1)(z-z_2)(z-z_3)(z-z_4)}\notag\\
\phi_4(z)&=\frac{\left[u_4\,(z-z_1)(z-z_2)z_{3 4} +\tfrac{1}{4} u_2^2\, \left((z-z_2)(z-z_3)z_{1 4}-(z-z_1)(z-z_4)z_{2 3}\right) \right]z_{1 2} z_{3 4}^2 {(d z)}^4}{{(z-z_1)}^2{(z-z_2)}^2{(z-z_3)}^3{(z-z_4)}^3}\notag\\
\phi_6(z)&=\frac{[u_6\, (z-z_3)(z-z_4)z_{12}+2\tilde{u}((z-z_1)(z-z_4)z_{23}\mp (z-z_2)(z-z_3)z_{14})]z_{12}{z_{34}}^4{(d z)}^6}{{(z-z_1)}^2{(z-z_2)}^2{(z-z_3)}^5{(z-z_4)}^5}\notag\\
\phi_8(z)&=\frac{u_8\, z_{1 2}^2 z_{3 4}^6 {(d z)}^8}{{(z-z_1)}^2{(z-z_2)}^2{(z-z_3)}^6{(z-z_4)}^6}\\
\phi_{10}(z)&=\frac{u_{10}\, {z_{12}}^2{z_{34}}^8{(d z)}^{10}}{{(z-z_1)}^2{(z-z_2)}^2{(z-z_3)}^8{(z-z_4)}^8}\notag\\
\tilde{\phi}(z)&=\frac{\tilde{u}\, z_{12}{z_{34}}^5{(d z)}^6}{{(z-z_1)}{(z-z_2)}{(z-z_3)}^5{(z-z_4)}^5}\quad.\notag
\end{align}
where the upper/lower sign in the expression for $\phi_6$ is for \eqref{SO126v2s}/\eqref{SO126v1s1sp}, respectively. The invariant $k$-differentials for \eqref{SO115v2s} are as above, but with $\tilde{u}\equiv0$.

\subsubsection{$Spin(12)+\tfrac{3}{2}(32)+4(12)$}\label{Spin12_c}
\begin{equation}
 \begin{matrix}\includegraphics[width=264pt]{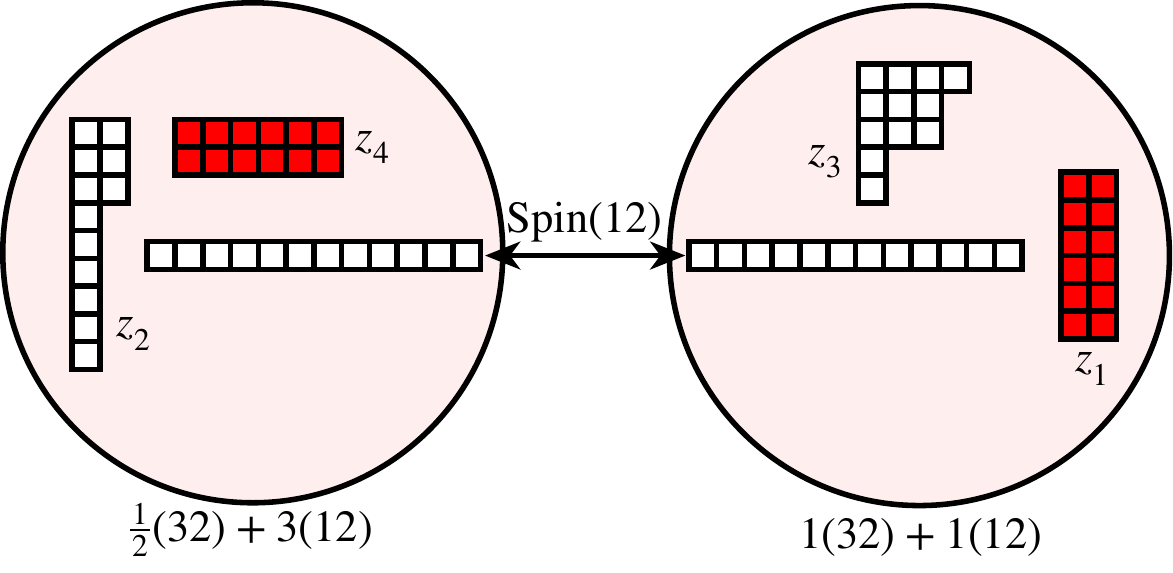}\end{matrix}\quad.
\label{SO124v3s}\end{equation}
The S-dual theories are an $SU(2)$ gauging of the $Sp(4)_{12} \times SU(2)_8 \times SU(2)_{128}$ SCFT
\begin{displaymath}
 \includegraphics[width=264pt]{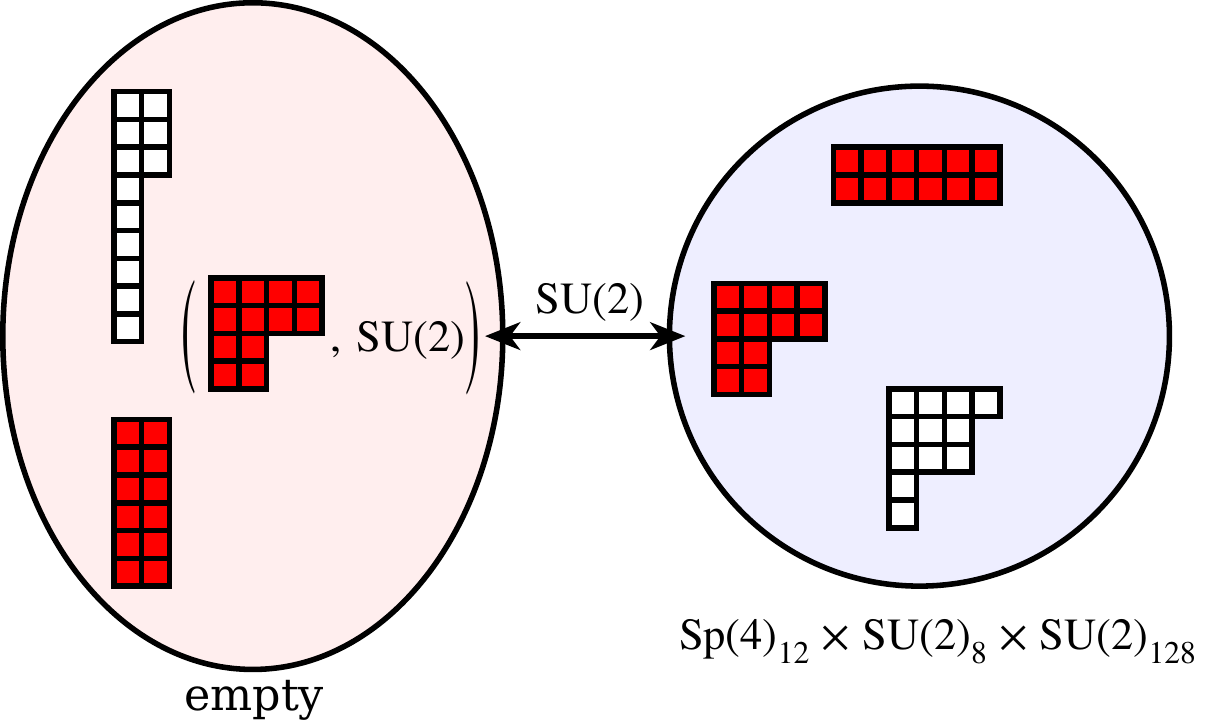}
\end{displaymath}
and a $G_2$ gauging of the $(F_4)_{16} \times Sp(4)_{12}$ SCFT

\begin{displaymath}
 \begin{matrix}\includegraphics[width=252pt]{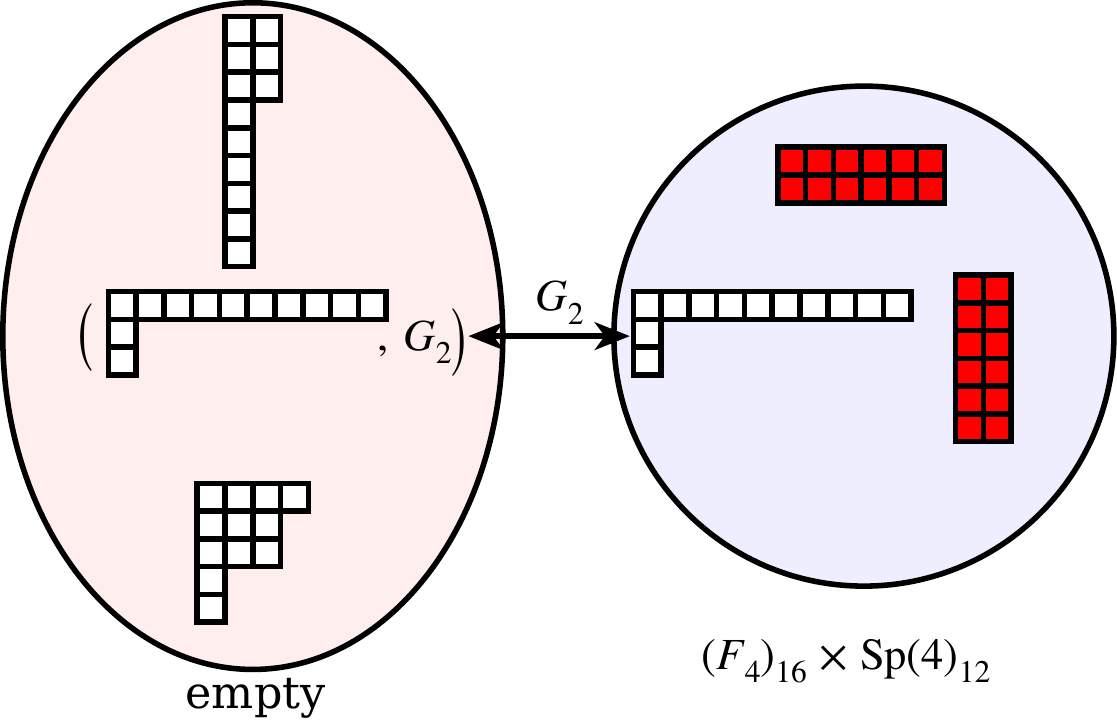}\end{matrix}\quad.
\end{displaymath}
The invariant $k$-differentials for \eqref{SO124v3s} are given by
\begin{align}\label{SO124v3sSol}
\phi_2(z)&=\frac{u_2 z_{12}z_{34} {(d z)}^2}{{(z-z_1)}{(z-z_2)}{(z-z_3)}{(z-z_4)}}\notag\\
\phi_4(z)&=\frac{[u_4 (z-z_2)z_{14}+\tfrac{1}{4}{u_2}^2(z-z_4)z_{12}]z_{12}{z_{34}}^2 {(d z)}^4}{{(z-z_1)}^2{(z-z_2)}^2{(z-z_3)}^2{(z-z_4)}^3}\notag\\
\phi_6(z)&=\frac{[u_6(z-z_1)(z-z_4)z_{23}-2\tilde{u}(z-z_1)(z-z_2)z_{34}+(2\tilde{u}+\tfrac{1}{4}u_2 u_4)(z-z_3)(z-z_4)z_{12}]z_{12}{z_{14}}{z_{34}}^3{(d z)}^6}{{(z-z_1)}^3{(z-z_2)}^2{(z-z_3)}^4{(z-z_4)}^5}\notag\\
\phi_8(z)&=\frac{[u_8(z-z_1)z_{34}+(\tfrac{1}{4}{u_4}^2+\tilde{u} u_2)(z-z_4)z_{13}]z_{14}{z_{12}}^2{z_{34}}^4{(d z)}^8}{{(z-z_1)}^4{(z-z_2)}^2{(z-z_3)}^5{(z-z_4)}^6}\\
\phi_10(z)&=\frac{[u_{10}(z-z_1)z_{34}+\tilde{u}u_4(z-z_4)z_{13}]{z_{12}}^2{z_{14}}^2{z_{34}}^5 {(d z)}^{10}}{{(z-z_1)}^5{(z-z_2)}^2{(z-z_3)}^6{(z-z_4)}^8}\notag\\
\tilde{\phi}(z)&=\frac{\tilde{u} z_{12}{z_{14}}^2{z_{34}}^3{(d z)}^6}{{(z-z_1)}^3{(z-z_2)}{(z-z_3)}^3{(z-z_4)}^5}\quad.\notag
\end{align}
For \eqref{SO113v3s}, they are as above, but with $\tilde{u}\equiv 0$.

\subsubsection{$Spin(12)+1(32)+\tfrac{1}{2}(32')+4(12)$}\label{Spin12_d}
\begin{equation}
 \begin{matrix}\includegraphics[width=264pt]{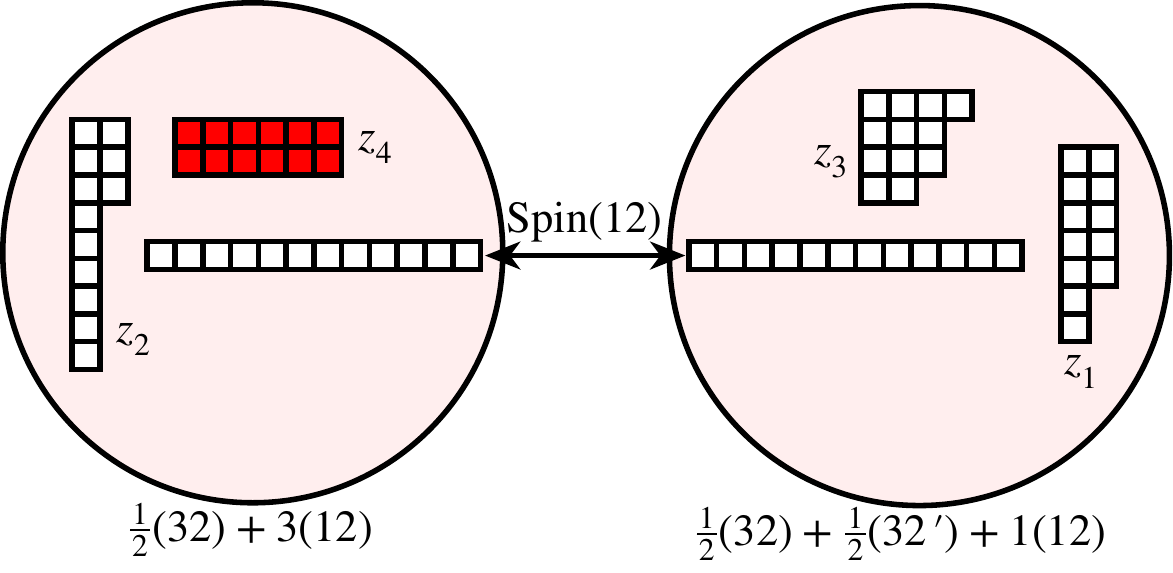}\end{matrix}\quad.
\label{SO124v2s1sp}\end{equation}
The S-dual theories are an $SU(2)$ gauging of the $Sp(4)_{12} \times SU(2)_8 \times U(1)$ SCFT
\begin{displaymath}
 \includegraphics[width=264pt]{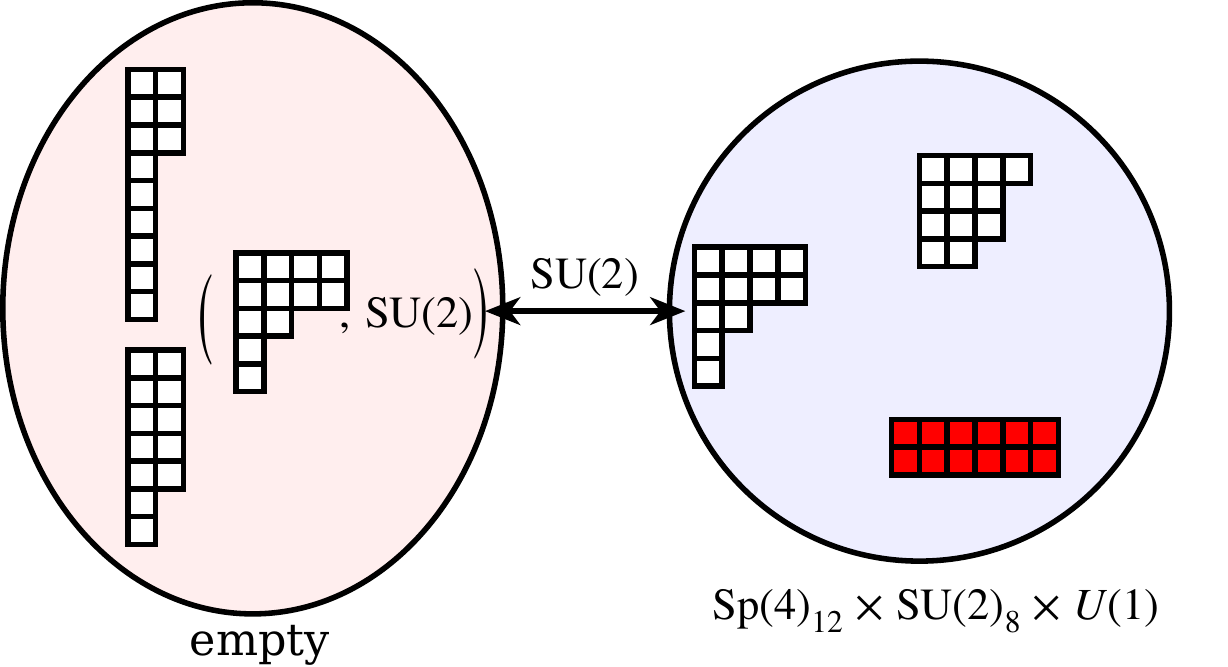}
\end{displaymath}
and a $G_2$ gauging of the $Sp(4)_{12} \times Spin(9)_{16}$ SCFT
\begin{displaymath}
 \begin{matrix}\includegraphics[width=274pt]{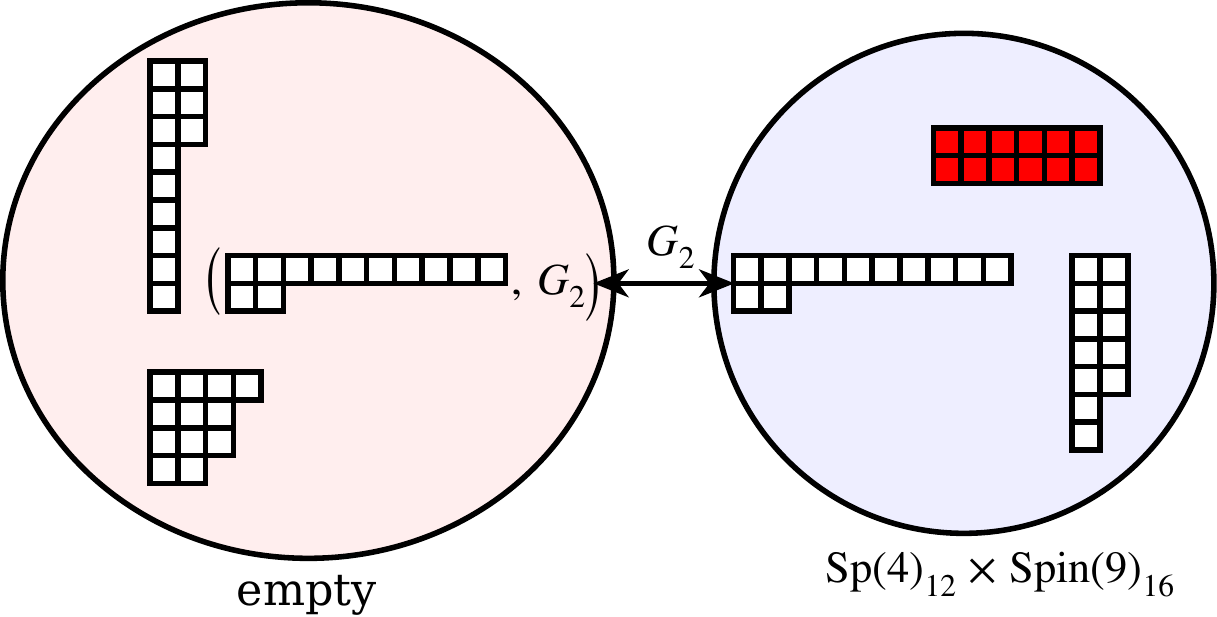}\end{matrix}\quad.
\end{displaymath}
The invariant $k$-differentials for \eqref{SO124v2s1sp} are given by
\begin{align}\label{SO124v2s1spSol}
\phi_2(z)&=\frac{u_2\, z_{1 2} z_{3 4} {(d z)}^2}{(z-z_1)(z-z_2)(z-z_3)(z-z_4)}\notag\\
\phi_4(z)&=\frac{\left[u_4\,(z-z_2)z_{1 4} +\tfrac{1}{4} u_2^2\, (z-z_4)z_{1 2} \right]z_{1 2} z_{3 4}^2 {(d z)}^4}{{(z-z_1)}^2{(z-z_2)}^2{(z-z_3)}^2{(z-z_4)}^3}\notag\\
\phi_6(z)&=\frac{[u_6(z-z_1)(z-z_4)z_{23}-2\tilde{u}(z-z_1)(z-z_2)z_{34}+\tfrac{1}{4}u_2u_4(z-z_3)(z-z_4)z_{12}]z_{12}z_{14}{z_{34}}^3 {(d z)}^6}{{(z-z_1)}^3{(z-z_2)}^2{(z-z_3)}^4{(z-z_4)}^5}\notag\\
\phi_8(z)&=\frac{\left[u_8\,(z-z_1)z_{3 4} +\tfrac{1}{4}u_4^2\, (z-z_4)z_{1 3} \right]z_{1 4} z_{1 2}^2 z_{3 4}^4 {(d z)}^8}{{(z-z_1)}^4{(z-z_2)}^2{(z-z_3)}^5{(z-z_4)}^6}\\
\phi_{10}(z)&=\frac{u_{10}{z_{1 2}}^2{z_{1 4}}^2{z_{3 4}}^6{(d z)}^{10}}{{(z-z_1)}^4{(z-z_2)}^2{(z-z_3)}^6{(z-z_4)}^8}\notag\\
\tilde{\phi}(z)&=\frac{\tilde{u}z_{12}z_{14}{z_{34}}^4{(d z)}^6}{{(z-z_1)}^2{(z-z_2)}{(z-z_3)}^4{(z-z_4)}^5}\quad.\notag
\end{align}
For \eqref{SO113v3s}, they are as above, but with $\tilde{u}\equiv 0$ (note that \eqref{SO124v2s1spSol} and \eqref{SO124v3sSol} become equal at $\tilde{u}=0$).

\subsubsection{$Spin(12)+2(32)+2(12)$}\label{Spin12_e}
\begin{equation}
 \begin{matrix}\includegraphics[width=264pt]{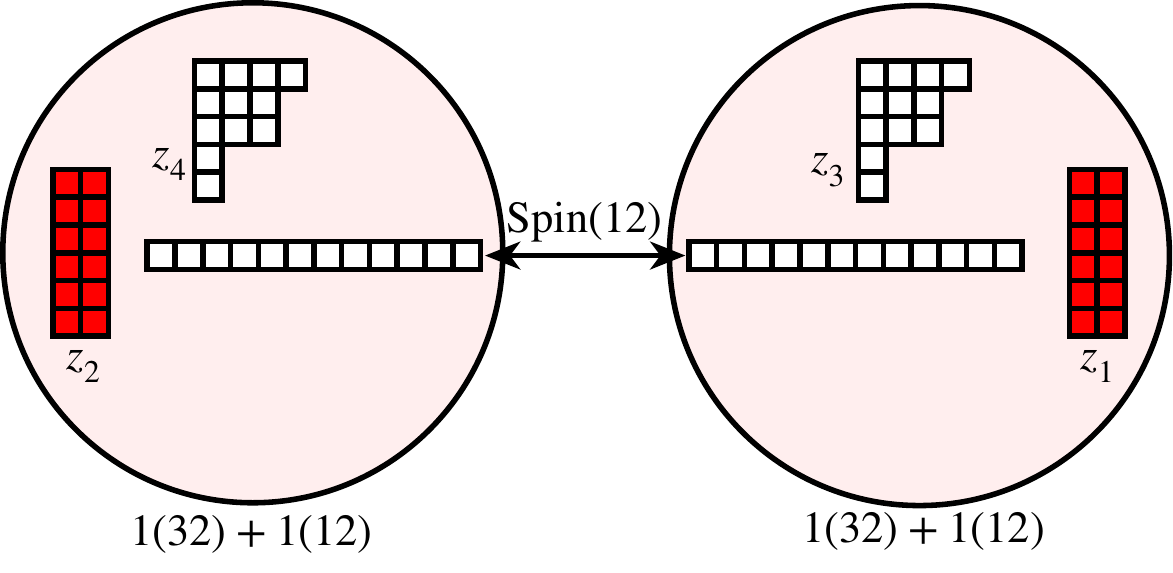}\end{matrix}\quad.
\label{SO122v4s}\end{equation}
The S-dual theory is an $Sp(3)$ gauging of the $Sp(3)_{11} \times {SU(2)}^2_{32}\, \text{SCFT} + \frac{5}{2}(6)$
\begin{displaymath}
 \begin{matrix}\includegraphics[width=264pt]{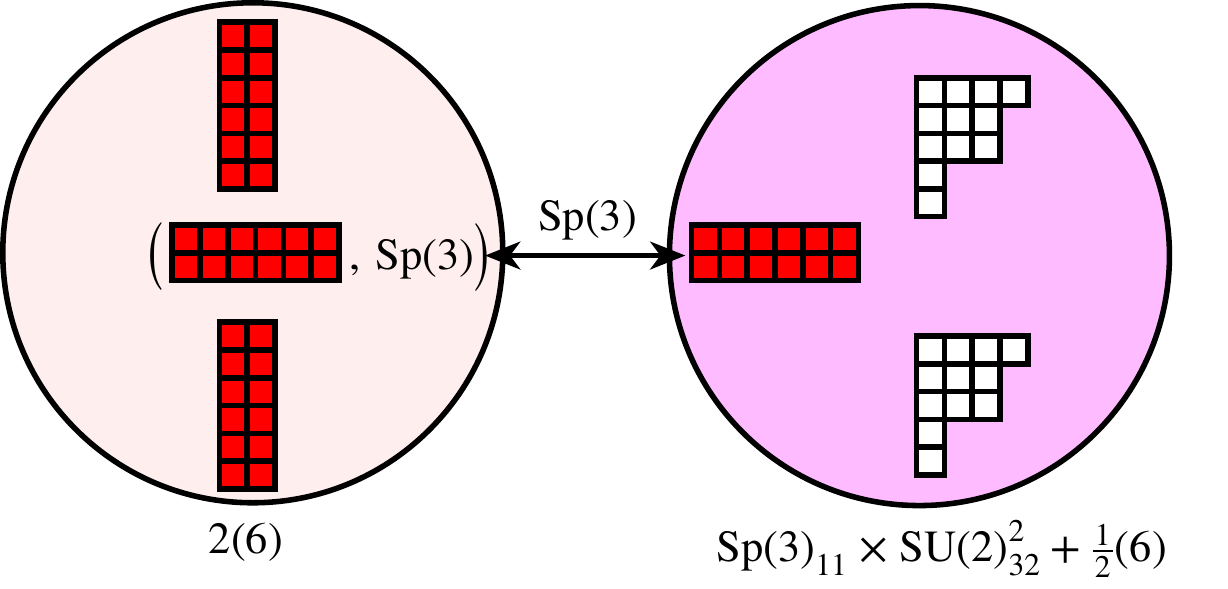}\end{matrix}\quad.
\end{displaymath}

The invariant $k$-differentials for \eqref{SO122v4s} are given by
\begin{align}\label{SO122v4sSol}
\phi_2(z)&=\frac{u_2z_{12}z_{34}{(d z)}^2}{{(z-z_1)}{(z-z_2)}{(z-z_3)}{(z-z_4)}}\notag\\
\phi_4(z)&=\frac{u_4{z_{12}}^2{z_{34}}^2{(d z)}^4}{{(z-z_1)}^2{(z-z_2)}^2{(z-z_3)}^2{(z-z_4)}^2}\notag\\
\phi_6(z)&=\frac{[u_6(z-z_1)(z-z_2)z_{34}-(2\tilde{u}+\tfrac{1}{2}u_2(u_4-\tfrac{1}{4}{u_2}^2))(z-z_1)(z-z_4)z_{23}}{{(z-z_1)}^3{(z-z_2)}^3{(z-z_3)}^4{(z-z_4)}^4}\notag\\
&\qquad\qquad\qquad\qquad \frac{+(2\tilde{u}+\tfrac{1}{2}u_2(u_4-\tfrac{1}{4}{u_2}^2))(z-z_2)(z-z_3)z_{14}]{z_{12}}^2{z_{34}}^3{(d z)}^6}{\vphantom{z^2}}\notag\\
\phi_8(z)&=\frac{[u_8(z-z_1)(z-z_2)z_{34}-(\tfrac{1}{4}{(u_4-\tfrac{1}{4}{u_2}^2)}^2+\tilde{u}u_2)(z-z_1)(z-z_4)z_{23}}{{(z-z_1)}^4{(z-z_2)}^4{(z-z_3)}^5{(z-z_4)}^5}\\
&\qquad\qquad\qquad\qquad \frac{+(\tfrac{1}{4}{(u_4-\tfrac{1}{4}{u_2}^2)}^2+\tilde{u}u_2)(z-z_2)(z-z_3)z_{14}]{z_{12}}^3{z_{34}}^4{(d z)}^8}{\vphantom{z^2}}\notag\\
\phi_{10}(z)&=\frac{[u_{10}(z-z_1)(z-z_2)z_{34}-\tilde{u}(u_4-\tfrac{1}{4}{u_2}^2)(z-z_1)(z-z_4)z_{23}}{{(z-z_1)}^5{(z-z_2)}^5{(z-z_3)}^6{(z-z_4)}^6}\notag\\
&\qquad\qquad\qquad\qquad \frac{+\tilde{u}(u_4-\tfrac{1}{4}{u_2}^2)(z-z_2)(z-z_3)z_{14}]{z_{12}}^4{z_{34}}^5{(d z)}^{10}}{\vphantom{z^2}}\notag\\
\tilde{\phi}(z)&=\frac{\tilde{u}{z_{12}}^3{z_{34}}^3{(d z)}^6}{{(z-z_1)}^3{(z-z_2)}^3{(z-z_3)}^3{(z-z_4)}^3}\quad.\notag
\end{align}

\subsubsection{$Spin(12)+\tfrac{3}{2}(32)+\tfrac{1}{2}(32')+2(12)$}\label{Spin12_f}
\begin{equation}
 \begin{matrix}\includegraphics[width=264pt]{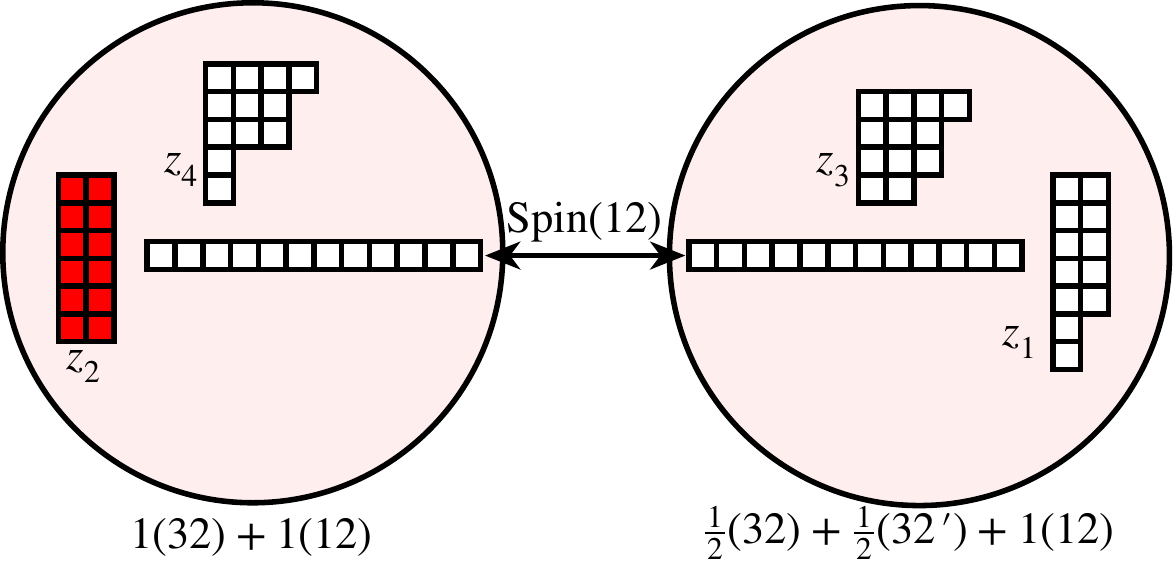}\end{matrix}\quad.
\label{SO122v3s1sp}\end{equation}
The S-dual theories are an $Sp(2)$ gauging of the $Sp(4)_{12} \times SU(2)_{128}$ SCFT
\begin{displaymath}
 \includegraphics[width=264pt]{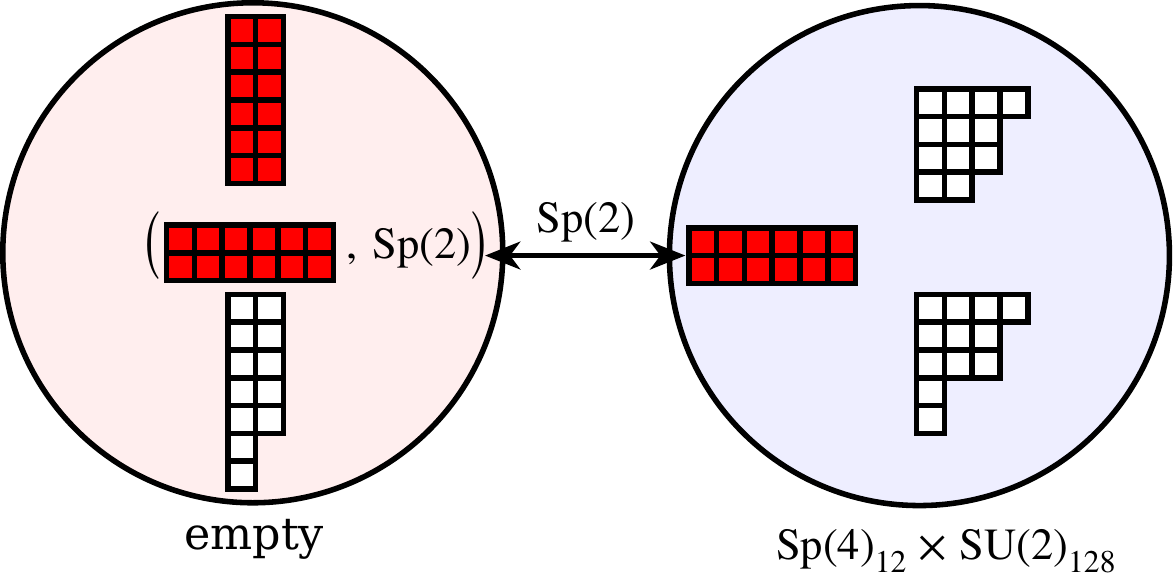}
\end{displaymath}
and a $Spin(11)$ gauging of the $(E_8)_{12}\, \text{SCFT} + \frac{3}{2}(32)$
\begin{displaymath}
 \begin{matrix}\includegraphics[width=289pt]{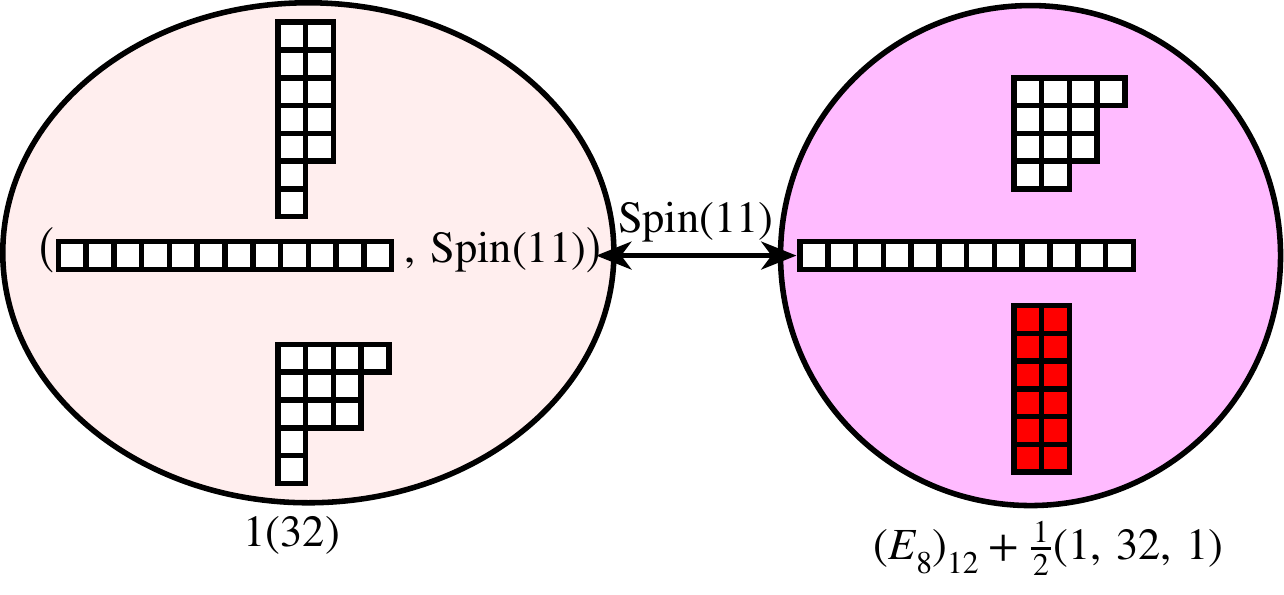}\end{matrix}\quad.
\end{displaymath}
The invariant $k$-differentials for \eqref{SO122v3s1sp} are given by
\begin{align}\label{SO122v3s1spSol}
\phi_2(z)&=\frac{u_2z_{12}z_{34}{(d z)}^2}{{(z-z_1)}{(z-z_2)}{(z-z_3)}{(z-z_4)}}\notag\\
\phi_4(z)&=\frac{u_4{z_{12}}^2{z_{34}}^2{(d z)}^4}{{(z-z_1)}^2{(z-z_2)}^2{(z-z_3)}^2{(z-z_4)}^2}\notag\\
\phi_6(z)&=\frac{
    \bigl[u_6(z-z_1)(z-z_2)z_{34}+(2\tilde{u}-\tfrac{1}{2}u_2(u_4-\tfrac{1}{4}{u_2}^2))(z-z_1)(z-z_4)z_{23}
  }{
    {(z-z_1)}^3{(z-z_2)}^3{(z-z_3)}^4{(z-z_4)}^4
  }\notag\\ &\qquad\qquad\qquad\qquad
  \frac{
    +\tfrac{1}{2}u_2(u_4-\tfrac{1}{4}{u_2}^2)(z-z_2)(z-z_3)z_{14}\bigr]{z_{12}}^2{z_{34}}^3{(d z)}^6
  }{\vphantom{z^2}}\notag\\
\phi_8(z)&=\frac{\bigl[u_8(z-z_1)(z-z_2)z_{34}-(\tfrac{1}{4}{(u_4-\tfrac{1}{4}{u_2}^2)}^2-\tilde{u}u_2)(z-z_1)(z-z_4)z_{23}}{{(z-z_1)}^4{(z-z_2)}^4{(z-z_3)}^5{(z-z_4)}^5}\\
&\qquad\qquad\qquad\qquad\frac{+\tfrac{1}{4}(u_4-\tfrac{1}{4}{u_2}^2)^2(z-z_2)(z-z_3)z_{14}\bigr]{z_{12}}^3{z_{34}}^4{(d z)}^8}{\vphantom{z^2}}\notag\\
\phi_{10}(z)&=\frac{[u_{10}(z-z_2)z_{34}+\tilde{u}(u_4-\tfrac{1}{4}u_2^2)(z-z_4)z_{23}]{z_{12}}^4{z_{34}}^5{(d z)}^{10}}{{(z-z_1)}^4{(z-z_2)}^5{(z-z_3)}^6{(z-z_4)}^6}\notag\\
\tilde{\phi}(z)&=\frac{\tilde{u}z_{23}{z_{12}}^2{z_{34}}^3{(d z)}^6}{{(z-z_1)}^2{(z-z_2)}^3{(z-z_3)}^4{(z-z_4)}^3}\quad.\notag
\end{align}

\subsubsection{$Spin(12)+1(32)+1(32')+2(12)$}\label{Spin12_f}
\begin{equation}
 \begin{matrix}\includegraphics[width=264pt]{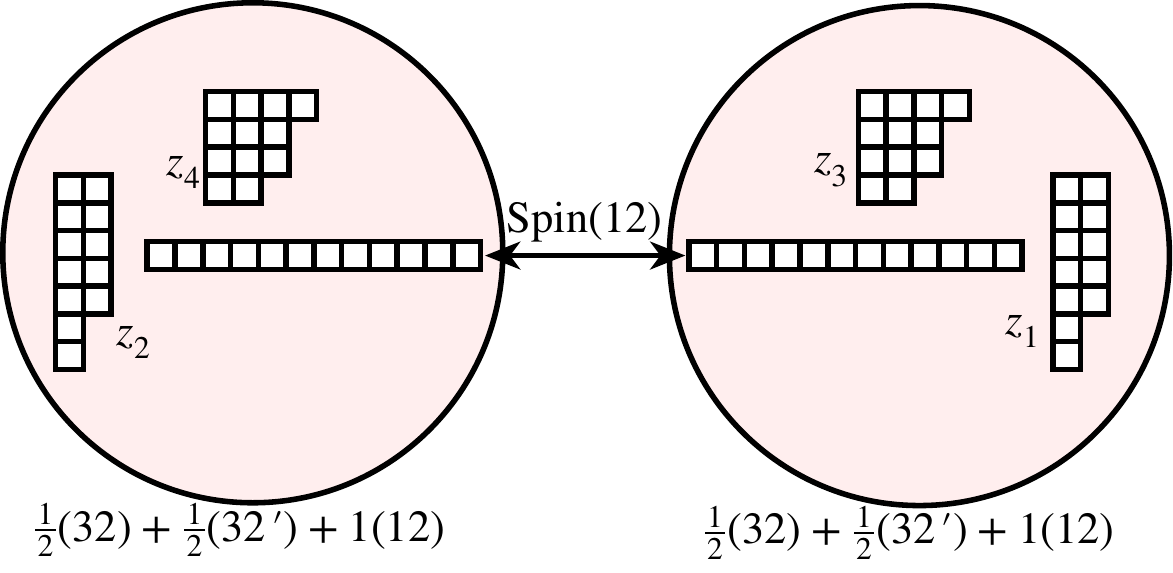}\end{matrix}\quad.
\label{SO122v2s2sp}\end{equation}
The S-dual theory is an $Sp(2)$ gauging of the $Sp(2)_{12} \times Sp(2)_{11} \times U(1)^2\,\text{SCFT}\, +\tfrac{1}{2}(4)$
\begin{displaymath}
 \begin{matrix}\includegraphics[width=289pt]{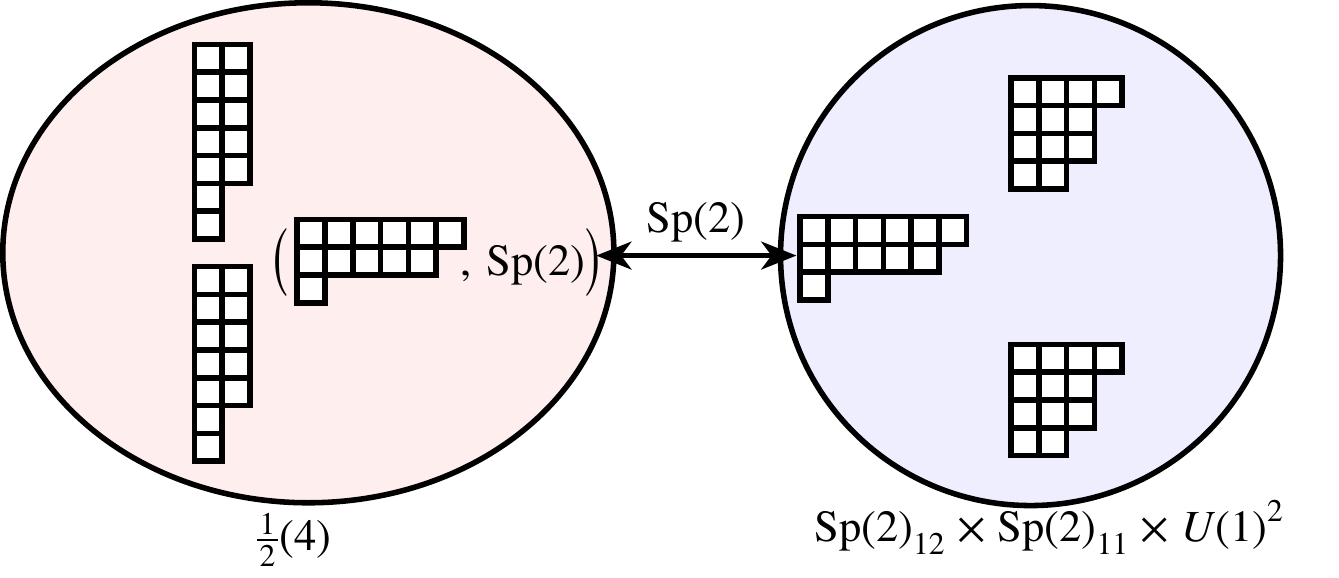}\end{matrix}.
\end{displaymath}
The invariant $k$-differentials for \eqref{SO122v2s2sp} are given by
\begin{align}\label{SO122v2s2spSol}
\phi_2(z)&=\frac{u_2\, z_{1 2} z_{3 4} {(d z)}^2}{(z-z_1)(z-z_2)(z-z_3)(z-z_4)}\notag\\
\phi_4(z)&=\frac{u_4\, z_{1 2}^2 z_{3 4}^2 {(d z)}^4}{{(z-z_1)}^2{(z-z_2)}^2{(z-z_3)}^2{(z-z_4)}^2}\notag\\
\phi_6(z)&=\frac{\left[u_6\,(z-z_1)(z-z_2)z_{3 4} -\tfrac{1}{2}u_2 \left(u_4-\tfrac{1}{4} u_2^2\right)\left((z-z_1)(z-z_3)z_{2 4}-(z-z_2)(z-z_4)z_{1 3}\right) \right]z_{1 2}^2 z_{3 4}^3 {(d z)}^6}{{(z-z_1)}^3{(z-z_2)}^3{(z-z_3)}^4{(z-z_4)}^4}\notag\\
\phi_8(z)&=\frac{\left[u_8\,(z-z_1)(z-z_2)z_{3 4} -\tfrac{1}{4}{\left(u_4-\tfrac{1}{4}u_2^2\right)}^2\left((z-z_1)(z-z_3)z_{2 4}-(z-z_2)(z-z_4)z_{1 3}\right)\right]z_{1 2}^3 z_{3 4}^4 {(d z)}^8}{{(z-z_1)}^4{(z-z_2)}^4{(z-z_3)}^5{(z-z_4)}^5}\notag\\
\phi_{10}(z)&=\frac{u_{10}{z_{12}}^4{z_{34}}^6{(d z)}^{10}}{{(z-z_1)}^4{(z-z_2)}^4{(z-z_3)}^6{(z-z_4)}^6}\\
\tilde{\phi}(z)&=\frac{\tilde{u}{z_{12}}^2{z_{34}}^4{(d z)}^6}{{(z-z_1)}^2{(z-z_2)}^2{(z-z_3)}^4{(z-z_4)}^4}\quad.\notag
\end{align}
For \eqref{SO111v4s}, they are as above, but with $\tilde{u}\equiv 0.$ As before, \eqref{SO122v4sSol},\eqref{SO122v3s1spSol} and \eqref{SO122v2s2spSol} become identical when you set $\tilde{u}=0$.

\subsubsection{More Spinors}\label{more_spinors}

We cannot obtain

\begin{itemize}%
\item $Spin(12) + \tfrac{5}{2}(32)$
\item $Spin(12) + 2(32) +\tfrac{1}{2}(32')$
\item $Spin(12) + \tfrac{3}{2}(32) +1(32')$

\end{itemize}
from compactifying the $D_6$ theory.

\section{$Spin(13)$ and $Spin(14)$ Gauge Theories}\label{Spin13_and_Spin14}
Here, we work in the $D_7$ theory.

\subsection{$Spin(13)+\tfrac{1}{2}(64)+7(13)$}\label{Spin13_a}
\begin{equation}
 \begin{matrix}\includegraphics[width=344pt]{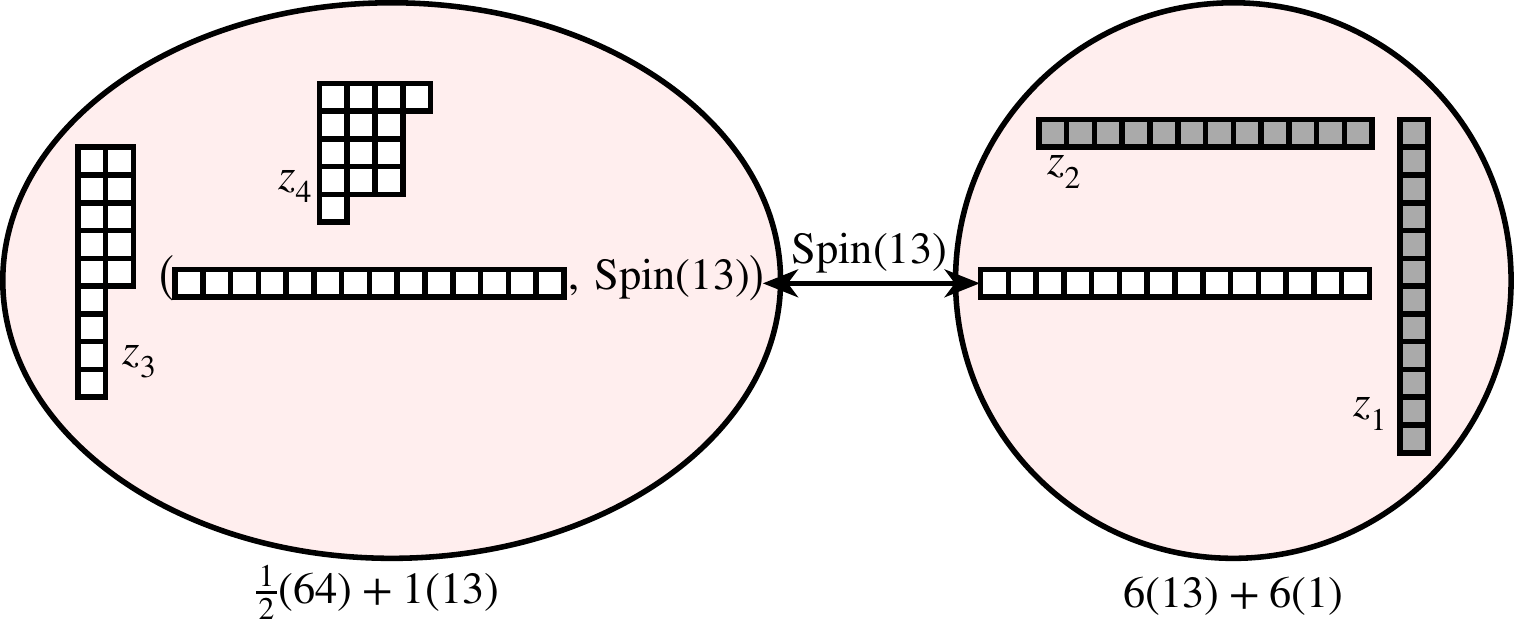}\end{matrix}\quad.
\label{SO137v1s}\end{equation}
Over the other degenerations, we have a gauge theory fixture

\begin{displaymath}
 \includegraphics[width=283pt]{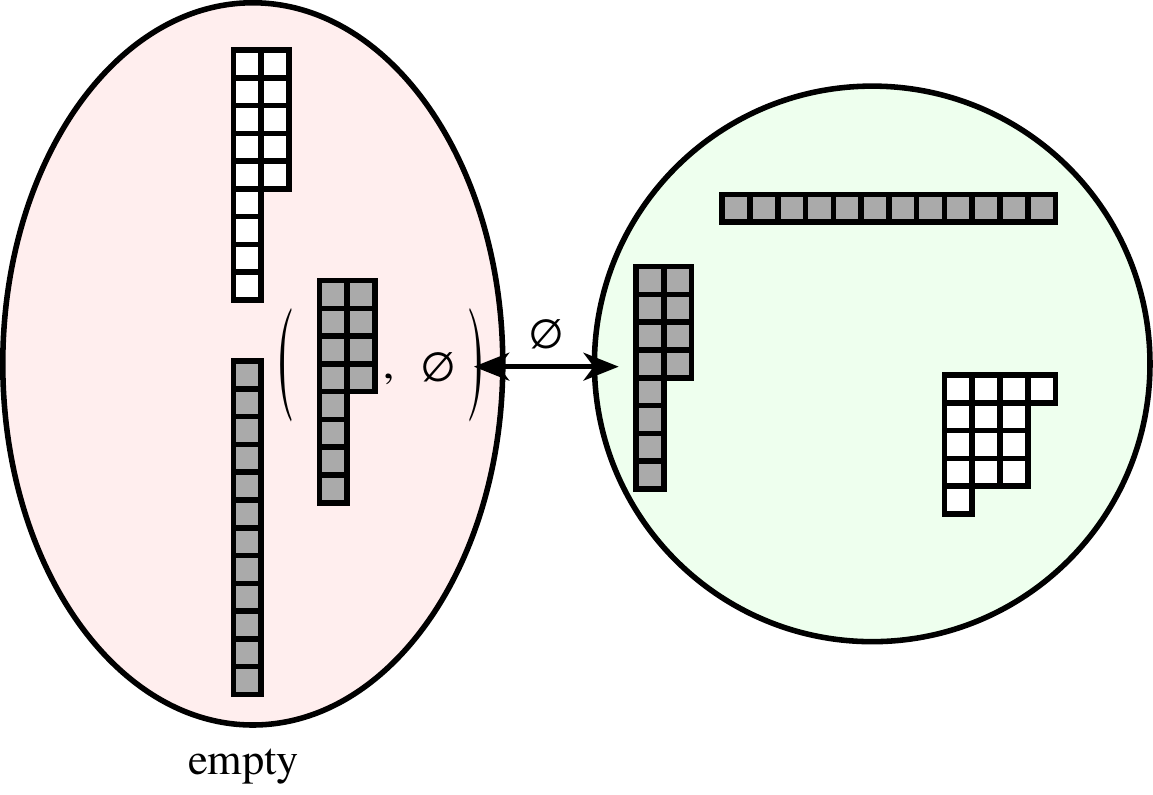}
\end{displaymath}
and an $SU(2)$ gauging of the $Sp(7)_{13} \times SU(2)_7 \text{SCFT} +\tfrac{1}{2}(2) + 6(1)$
\begin{displaymath}
 \begin{matrix}\includegraphics[width=316pt]{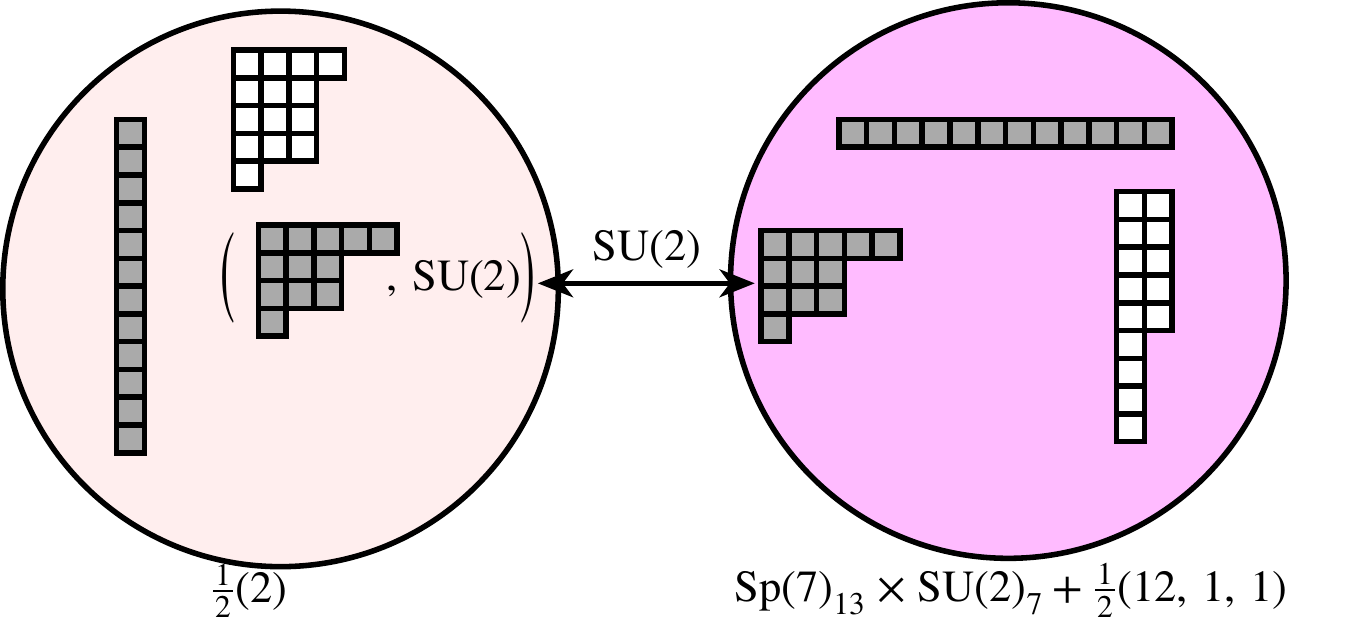}\end{matrix}.
\end{displaymath}
The invariant $k$-differentials for \eqref{SO137v1s} are given by
\begin{align}\label{Spin13sol}
\phi_2(z)&=\frac{u_2\,z_{13}z_{24}{(d z)}^2}{{(z-z_1)}{(z-z_2)}{(z-z_3)}{(z-z_4)}}\notag\\
\phi_4(z)&=\frac{u_4\,z_{13}z_{23}{z_{24}}^2{(d z)}^4}{{(z-z_1)}{(z-z_2)}^3{(z-z_3)}^2{(z-z_4)}^2}\notag\\
\phi_6(z)&=\frac{[u_6(z-z_3)z_{12}-\tfrac{1}{2}u_2(u_4-\tfrac{1}{4}{u_2}^2)(z-z_2)z_{13}]z_{23}z_{34}{z_{24}}^3{(d z)}^6}{{(z-z_1)}{(z-z_2)}^5{(z-z_3)}^3{(z-z_4)}^4}\notag\\
\phi_8(z)&=\frac{[u_8(z-z_3)z_{12}-\tfrac{1}{4}(u_4-\tfrac{1}{4}{u_2}^2)^2(z-z_2)z_{13}]z_{34}{z_{23}}^2{z_{24}}^4{(d z)}^8}{{(z-z_1)}{(z-z_2)}^7{(z-z_3)}^4{(z-z_4)}^5}\\
\phi_{10}(z)&=\frac{u_{10}\,z_{13}z_{24}{z_{23}}^3{z_{24}}^5{(d z)}^{10}}{{(z-z_1)}{(z-z_2)}^9{(z-z_3)}^4{(z-z_4)}^6}\notag\\
\phi_{12}(z)&=\frac{u_{12}\,z_{13}z_{24}{z_{23}}^3{z_{24}}^7{(d z)}^{12}}{{(z-z_1)}{(z-z_2)}^{11}{(z-z_3)}^4{(z-z_4)}^8}\notag\\
\noalign{\text{and}}
\tilde{\phi}(z)&=0\quad.\notag
\end{align}

\subsection{More Spinors}\label{give_up}
We cannot obtain

\begin{itemize}%
\item $Spin(13)+1(64)+3(13)$
\item $Spin(14)+1(64)+4(14)$
\end{itemize}

\noindent
from compactifying the $D_7$ theory.

\section{Higher $N$?}\label{no_help}
For the ``missing" theories of \S\ref{more_spinors} and \S\ref{give_up}, we might hope to find realizations in the higher $D_N$ or $A_{2N-1}$ theories. It is easy to see that is no help. The key realization is that we need a candidate free-field fixture, consisting of three regular punctures. One of these punctures must be a full puncture.

In the $D_N$ theory, the full puncture, $[1^{2N}]$, has a $Spin(2N)_{4(N-1)}$ flavour symmetry. The free fields transform as some representation of $Spin(2N)$ which reproduce the level $k=4(N-1)$. If the representation should \emph{happen} to decompose correctly under a $Spin(12)$ (\emph{mutatis mutandis} for a $Spin(13)$ or $Spin(14)$) subgroup, then we would have a chance to build a realization of one of our missing gauge theories.
\begin{itemize}%
\item For the $Spin(12)$ theories of \S\ref{more_spinors}, we could note that the $64$ of $Spin(14)$ decomposes as $1(32)+1(32')$. But getting the right level would require a puncture with level $k=32$, whereas the full puncture of the $D_7$ theory has only $k=24$.
\item For the $Spin(13)$ and $Spin(14)$ theories of \S\ref{give_up}, going to higher $D_N$ could only produce the $64$ with multiplicity $>1$, which also does not help.
\end{itemize}

In the twisted sector of the $A_{2N-1}$ theory, the full puncture has $Spin(2N+1)_{2(2N-1)}$ flavour symmetry.
\begin{itemize}%
\item For the $Spin(12)$ theories of \S\ref{more_spinors}, we need $k$ to be a multiple of $8$, so none of these are satisfactory.
\item For the $Spin(13)$ and $Spin(14)$ theories of \S\ref{give_up}, we need $k$ to be a multiple of $4$, which also does not work.
\end{itemize}

What about the exceptional (2,0) theories? $E_7$ and $E_8$ contain our desired gauge groups as subgroups. But neither the $56$ of $E_7$, nor the $248$ of $E_8$ decompose correctly to provide candidate free field fixtures with one full puncture (and two other regular punctures).
   
So it appears that the missing theories of \S\ref{more_spinors} and \S\ref{give_up}, are not realizable as compactifications of the $(2,0)$ theory.

\section*{Acknowledgements}\label{Acknowledgements}
\addcontentsline{toc}{section}{Acknowledgements}

We would like to thank D.~Ben-Zvi, S.~Katz, G.~Moore, D.~Morrison, A.~Neitzke, R.~Plesser and Y.~Tachikawa for helpful discussions. J.~D.~and O.~C.~would like to thank the Aspen Center for Physics (supported, in part, by the National Science Foundation under Grant PHY-1066293) for their hospitality when this work was initiated. O.~C.~would further like to thank the Simons Foundation for partial support in Aspen. The work of J.~D.~and A.~T.~was supported in part by the National Science Foundation under Grant PHY-1316033. The work of O.~C.~was supported in part by the INCT-Matem\'atica and the ICTP-SAIFR in Brazil through a Capes postdoctoral fellowship.

\begin{appendices}

\end{appendices}

\bibliographystyle{utphys}
\bibliography{ref}

\end{document}